\documentclass[prd,preprintnumers,superscriptaddress,nofootinbib,showpacs,twocolumn]{revtex4-1}
\pdfoutput=1
\usepackage{pstricks}
\usepackage{color}
\usepackage{amssymb,amsmath,bm,bbm}
\usepackage{epsf,epsfig}
\usepackage{afterpage}
\usepackage{longtable}
\usepackage{latexsym,mathrsfs,dsfont}
\usepackage{graphics}
\usepackage{url}
\usepackage{paralist}
\usepackage{bbold}
\usepackage{appendix}

\newcommand{\be}{\begin{equation}}
\newcommand{\ee}{\end{equation}}
\newcommand{\bi}{\begin{itemize}}
\newcommand{\ei}{\end{itemize}}
\newcommand{\ba}{\begin{array}}
\newcommand{\ea}{\end{array}}
\newcommand{\bea}{\begin{eqnarray}}
\newcommand{\eea}{\end{eqnarray}}
\newcommand{\bec}{\begin{center}}
\newcommand{\eec}{\end{center}}

\newcommand{\nn}{\nonumber}

\newcommand{\gmn}{g_{M N}}% metric tensor
\def\@seccntformat#1{\@ifundefined{#1@cntformat}%
   {\csname the#1\endcsname\quad}%      default
   {\csname #1@cntformat\endcsname}%    enable individual control
}

\begin{document}

\preprint{BARI-TH/20-725}

\title{Chaos in a $Q \bar Q$ system at finite temperature and baryon density}
\author{P.~Colangelo}
\affiliation{Istituto Nazionale di Fisica Nucleare, Sezione di Bari, via Orabona 4, 70126 Bari, Italy}
\author{F.~De~Fazio}
\affiliation{Istituto Nazionale di Fisica Nucleare, Sezione di Bari, via Orabona 4, 70126 Bari, Italy}
\author{N.~Losacco}
\affiliation{Dipartimento Interateneo di Fisica "Michelangelo Merlin", Universit\`a degli Studi di Bari, via Orabona 4, 70126 Bari, Italy}
\affiliation{Istituto Nazionale di Fisica Nucleare, Sezione di Bari, via Orabona 4, 70126 Bari, Italy}

\begin{abstract}
\noindent
Onset of  chaos  for the holographic dual of a  $Q \bar Q$ system at finite temperature and baryon density is studied.  We consider a string in the AdS Reissner-Nordstrom  background near the black-hole horizon, and investigate   small time-dependent perturbations of the static configurations.  The proximity to the horizon induces  chaos, which is softened increasing the chemical potential. A background geometry  including the effect of a dilaton is also examined. The  Maldacena, Shenker and Stanford bound on the Lyapunov  exponents characterizing the perturbations is satisfied for finite baryon chemical potential and when the dilaton is included in the metric.
\end{abstract}

\vspace*{2cm}
%\pacs{12.60.-i , 13.25.Hw}

\maketitle

\section{Introduction}

It has been recently conjectured under general assumptions that, for a  thermal quantum system  at temperature $T$, some out-of-time-ordered correlation functions involving  Hermitian operators, for determined time intervals, have an exponential  time dependence characterized by an exponent  $\lambda$, and that such exponent obeys the bound 
\begin{equation}
\lambda \le 2 \pi T
\label{eq:1}
\end{equation}
(in units in which  $\hbar=1$ and $k_B=1$).
The correlation functions are related to the thermal expectation values of the (square) commutator of two Hermitian operators at a time separation $t$, which quantify the effect
of one operator on later measurements of the other one, a framework for introducing chaos for a quantum system.
The  conjectured bound, proposed by Maldacena, Shenker and Stanford \cite{Maldacena:2015waa}, is remarkable due to its generality. It has been inspired by the observation that black holes (BH)
are the fastest `scramblers'' in nature: the time needed for a system near a BH horizon to loose  information depends  logarithmically on the number of degrees of freedom of the system \cite{Sekino_2008, susskind2011addendum}. The consequences on the connection between chaotic quantum systems and gravity  have been soon investigated   \cite{Shenker_2014,Shenker:2014cwa,Kitaev,Polchinski:2015cea}. A relation between the size of operators on the boundary quantum theory, involved in the temporal evolution of a perturbation, and the momentum of a particle falling in the bulk has  been  proposed in a holographic framework \cite{Susskind:2018tei,Brown:2018kvn}.

A generalization of the bound \eqref{eq:1}  for a thermal quantum system with  a global symmetry has been proposed \cite{Halder:2019ric}:
\begin{equation}
\lambda \le \frac{2 \pi T}{1- \big|\frac{\mu}{\mu_c}\big|},
\label{eq:2}
\end{equation}
where $\mu$ is the chemical potential  related to the global symmetry, and  $\mu_c$  is a critical value above which the thermodynamical ensemble is not defined. The inequality \eqref{eq:2} is conjectured for $\mu \ll \mu_c$ and relaxes the bound \eqref{eq:1}. Our purpose is to test this generalization.

Several analyses have been devoted to check Eq.~\eqref{eq:1} using the  AdS/CFT correspondence \cite{Maldacena_1998, Witten:1998qj, Gubser_1998}, adopting a dual geometry with a black hole, identifying  $T$  with  the  Hawking temperature,  for example in \cite{deBoer:2017xdk,Dalui:2018qqv}. In particular, the heavy quark-antiquark pair, described holographically by a string hanging in the bulk  with end points on the boundary \cite{Avramis:2006nv,Avramis:2007mv,Arias_2010,Nunez:2009da}, has been studied in this context    \cite{Hashimoto:2018fkb, Ishii_2017,Akutagawa_2019}. For this system $\lambda$ is  the Lyapunov exponent characterizing the chaotic behavior of time-dependent fluctuations around the static  configuration.

To test the generalized bound \eqref{eq:2} one has to include  the chemical potential in the holographic description. In QCD, a $U(1)$ global symmetry is connected to the conservation of the baryon number.  A dual metric has been identified with the AdS Reissner-Nordstrom  (RN)  metric for a charged
 black hole. We can use such a background  for testing Eq.~\eqref{eq:2}.

The discussion of the $5d$  AdS-RN metric as a dual  geometry for a thermal system with conserved baryon number can be found, e.g., in  \cite{Lee_2009,Colangelo:2010pe}.
The  metric  is defined by the line element
\begin{equation}
d s^2 = -r^2 f \left( r \right) d t^2 + r^2 d {\bar x}^2 +\frac{1}{r^2 f \left( r \right)} d r^2,
\label{eq:21}
\end{equation}
with $r$ the radial bulk coordinate and
\be
f \left( r \right)=1-\frac{r_H^4}{r^4} -\frac{\mu^2 r_H^2}{r^4} + \frac{\mu^2 r_H^4}{r^6}.
\label{eq:20}
\ee
The geometry has an  outer  horizon  located at $r=r_H$, and the Hawking temperature is 
\be
T_H =\frac{r_H}{\pi} \left( 1- \frac{\mu^2}{2 r_H^2} \right) \,. \label{eq:thBH}
\ee
The constant $\mu \le {\sqrt 2} \, r_H$ in \eqref{eq:20} is interpreted as the baryon chemical potential of the boundary theory, and is holographically related to the charge $\hat Q$ of the RN black hole: $\hat Q=\mu^2/r_H^2$.

\begin{figure}[t!]
	\centering
	\includegraphics[width=0.35\textwidth]{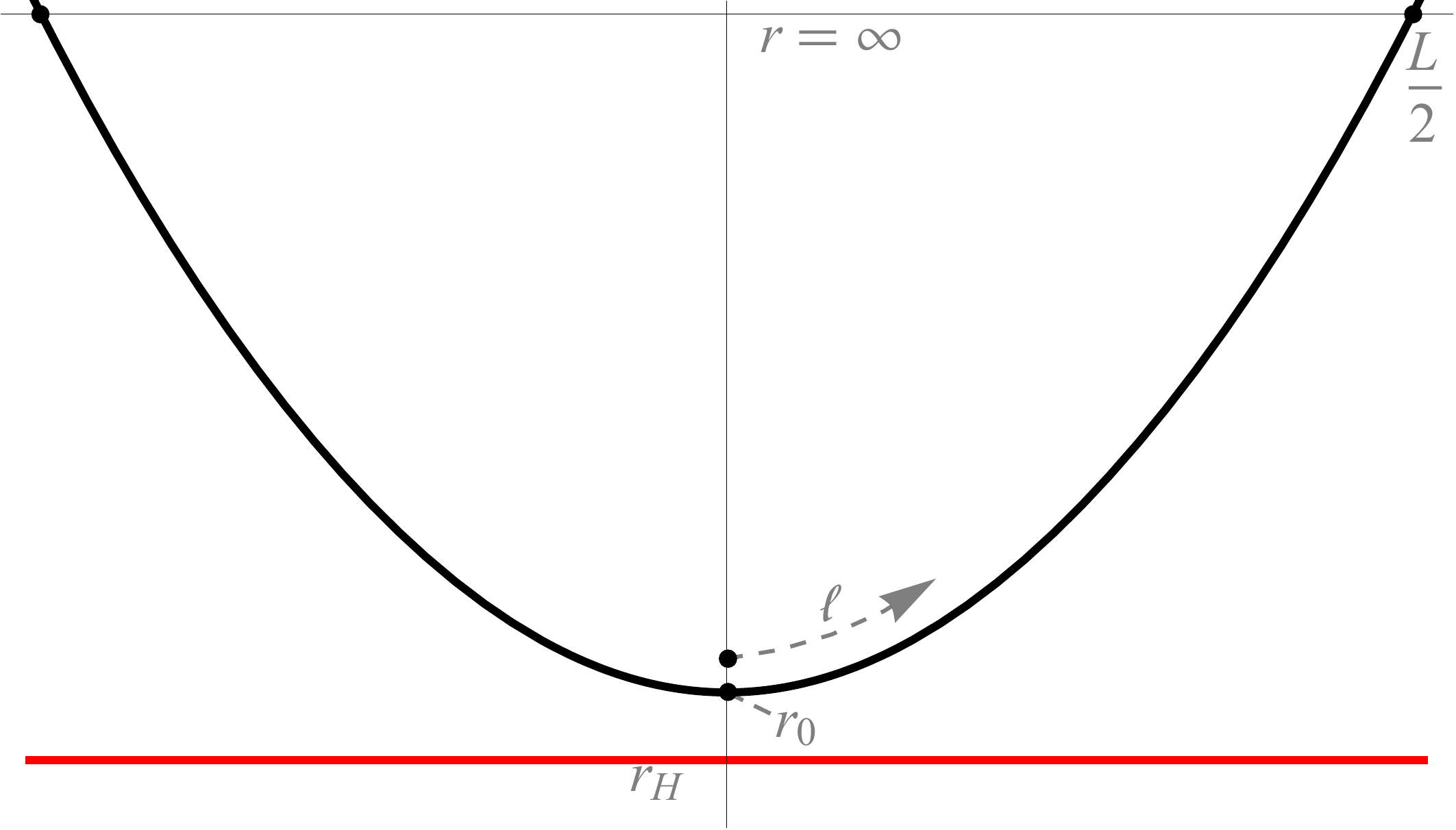}
	\caption[String profile]{\baselineskip 12 pt \small Profile of the static string for the $Q \bar Q$ system. $r_0$ is the position of the tip of the string, $r_H$  the position of the  horizon,  $L$  the distance between the end points on the boundary.}
	\label{Fig:1}
\end{figure}

The gravity dual of the heavy quark $Q \bar Q$ system at finite temperature and chemical potential  is a string in the background  \eqref{eq:21},\eqref{eq:20}  with the endpoints on the boundary (Fig.~\ref{Fig:1}).  To investigate the onset of chaos for this system
focusing on the effects of the chemical potential, we use the same approach adopted in \cite{Hashimoto:2018fkb} for the system at $\mu=0$, to shed light on the differences 
with respect to the case of  vanishing chemical potential. 

\section{Generalities of the suspended string in a gravity background}
The AdS-RN metric in (\ref{eq:21}) belongs to a general class of geometries described by the line element
\be
ds^2=-g_{tt}(r) dt^2+g_{xx}(r)d{\bar x}^2 +g_{rr}(r)dr^2 \,\, . \label{general}
\ee
The dynamics of a string in such a background is governed by the Nambu-Goto (NG) action
\begin{equation}
S=-\frac{1}{2 \pi \alpha^\prime} \int \mathrm{d}\tau \mathrm{d}\sigma \sqrt{\det \Big(\gmn \partial_a X^{M} \partial_b X^{N}\Big)}, \;\;\; 
\label{eq:22}
\end{equation}
\noindent
with $a,b=(\tau,\sigma)$ and  $\alpha^\prime$    the  string tension.
$\gmn$ is the metric tensor in  \eqref{general}, $X^{M}$ are the $5d$ coordinates and  the derivatives are with respect to  the world sheet coordinates $\tau$ and $\sigma$. 

We denote  by $r_0$  the position of the tip of the string  as shown in Fig.~\ref{Fig:1}, and $l$
  the proper distance   measured along the string starting from $r_0$. Choosing  $\tau = t$ and $\sigma =l$ ($l$-gauge), for a static string laying in the $x-r$ plane with $X^{M}= \left(t,x( l ),0,0,r(l) \right)$  the Nambu-Goto action  reads:
\begin{equation}
S=- \frac{\mathcal{T}}{2 \pi \alpha^\prime} \int_{-\infty}^{\infty} \mathrm{d}l \sqrt{F^2  \left( r \right) \acute{x}^2 \left( l \right) + G^2  \left( r \right) \acute{r}^2 \left( l \right) },
\label{eq:42}
\end{equation}
\noindent 
where $\acute{x}=\frac{\partial x}{\partial l}$ and $\acute{r}=\frac{\partial r}{\partial l}$,  $F^2 \left( r \right)= g_{tt} \left( r \right) g_{xx} \left( r \right)$ and $G^2 \left( r \right)= g_{tt} \left( r \right) g_{rr} \left( r \right)$. 
For the metric  \eqref{eq:21} one has  $F^2(r)=r^4 f(r)$ and $G(r)=1$.

 Note that $x$ is a cyclic coordinate, hence:

\begin{equation}
\acute{x} \left( l \right) = \pm \frac{\acute{r}\left( l \right)}{\sqrt{\frac{r^4 f \left( r \right)}{r_0^4 f \left( r_0 \right)} \left( r^4 f \left( r \right) -r_0^4 f \left( r_0 \right)  \right)}} .
\label{eq:43}
\end{equation} 

The  solution of Eq.~\eqref{eq:43} is obtained  considering that

\begin{equation}
\mathrm{d}l^2=g_{xx} \left( r \right) \mathrm{d}x^2 + g_{rr} \left( r \right) \mathrm{d}r^2.
\label{eq:44}
\end{equation}
\noindent 
For the unit vector $t^M = \left( 0, \acute{x} \left( l \right),0,0,\acute{r} \left( l \right) \right)$ tangent  to the string at the point with coordinate $l$  the relation holds:

\begin{equation}
\begin{aligned}
\gmn t^M t^N &=g_{xx} \left( r \right) \acute{x}^2 \left( l \right) + g_{rr} \left( r \right) \acute{r}^2 \left( l \right)\\
&= r^2  \acute{x}^2 \left( l \right) + \frac{1}{r^2 f \left( r \right)}  \acute{r}^2 \left( l \right)=1.
\end{aligned}
\label{eq:45}
\end{equation}
\noindent 
Including this constraint in Eq.~(\ref{eq:43}) gives

\bea
\acute{r}&=& \pm \frac{\sqrt{r^4 f \left( r \right)-r_0^4 f \left( r_0 \right)}}{r}
\label{eq:46} \\
\acute{x}&=& \pm \frac{\sqrt{r_0^4 f \left( r_0 \right)}}{r^3 \sqrt{f \left( r \right)}}.
\label{eq:47}
\eea
\noindent 
The function  $r \left( l \right)$  for the static string can be computed integrating Eq.~(\ref{eq:46}).

\begin{figure}[b!]
	\centering
	\includegraphics[width=0.47\textwidth]{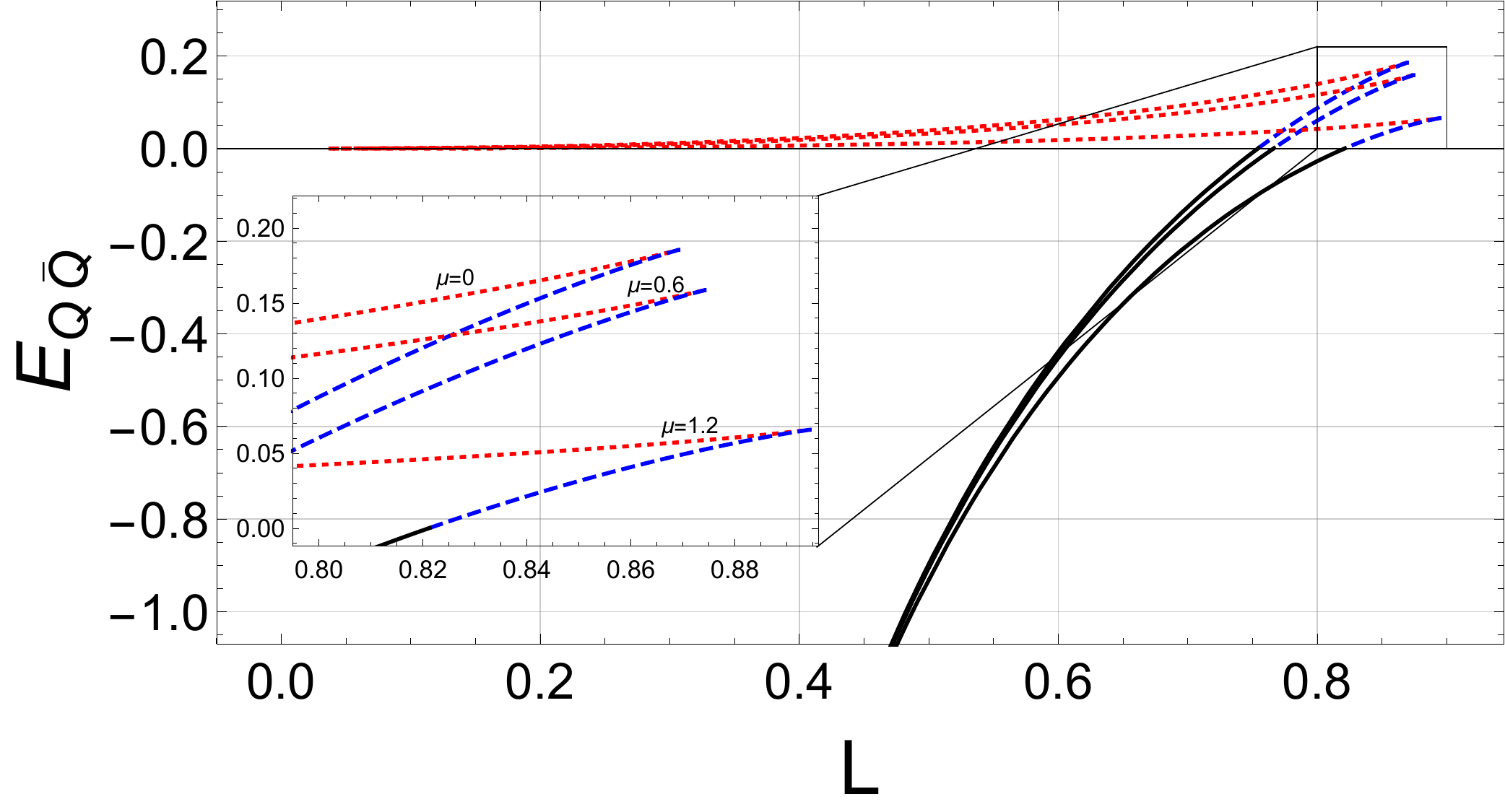}
	\caption{\baselineskip 12 pt \small  Double valued $E_{Q{\bar Q}}(L)$ for $r_H=1$ and $\mu=0,\,0.6,\,1.2$. The inset is an enlargement of  the $L \simeq 0.8$ range.}
	\label{fig:EvsL}
\end{figure}

The dependence of $L$,  the distance  between the string endpoints on the boundary, on $r_0$ is obtained:
\be
L(r_0)=2 \int_{r_0}^\infty dr \, \frac{r_0^2 \sqrt{f(r_0)}}{r^2 \sqrt{f(r)}\sqrt{r^4 f(r)-r_0^4 f(r_0)}} \,.
\label{Lvsr0}
\ee
The energy of the string configuration 
\be
E(r_0)= \frac{1}{\pi \alpha^\prime} \int_{r_0}^\infty dr  \, \frac{r^2 \sqrt{f(r)}}{\sqrt{r^4 f(r)-r_0^4 f(r_0)}} \,\,
\label{Evsr0}
\ee
diverges and needs to be regularized. A possible prescription is to subtract the bare quark masses, interpreted as the energy of the string consisting in two straight lines from the boundary  to the horizon,
\be
m_Q=\frac{1}{2 \pi \alpha^\prime} \int_{r_{H}}^\infty dr \,\, ,
\ee
obtaining
\bea
&&
E_{Q{\bar Q}}(r_0)= \nn \\
&&\frac{1}{\pi \alpha^\prime}
 \left( \int_{r_0}^\infty dr  \, \frac{r^2 \sqrt{f(r)}}{\sqrt{r^4 f(r)-r_0^4 f(r_0)}}- \int_{r_H}^\infty dr  \right) \, . \nn \\
\eea 
The function $E_{Q{\bar Q}}$ can be expressed vs $L$.
For the  metric in Eq.~\eqref{eq:20}, the distance   $L(r_0)$   has a  maximum  $L_{max}$, and all  values   $L \in [0, \, L_{max}]$ are obtained for two positions $r_0$. Also the function $E_{Q{\bar Q}}(r_0)$ has a maximum, which decreases and is reached earlier as $\mu$ increases.
For each value of  the chemical potential there is a value of $r_0$ above which there is one  energy value indicating a stable string configuration. Below such $r_0$, as shown in  
Fig.~\ref{fig:EvsL}, the  $E_{Q \bar Q}(L)$ is not single valued:  for each   $L$ there are  profiles identified by different  $r_0$, with different energies,  corresponding to stable and  unstable configurations.

\section{Square string}\label{square}
\begin{figure}[b!]
	\centering
	\includegraphics[width=0.4 \textwidth]{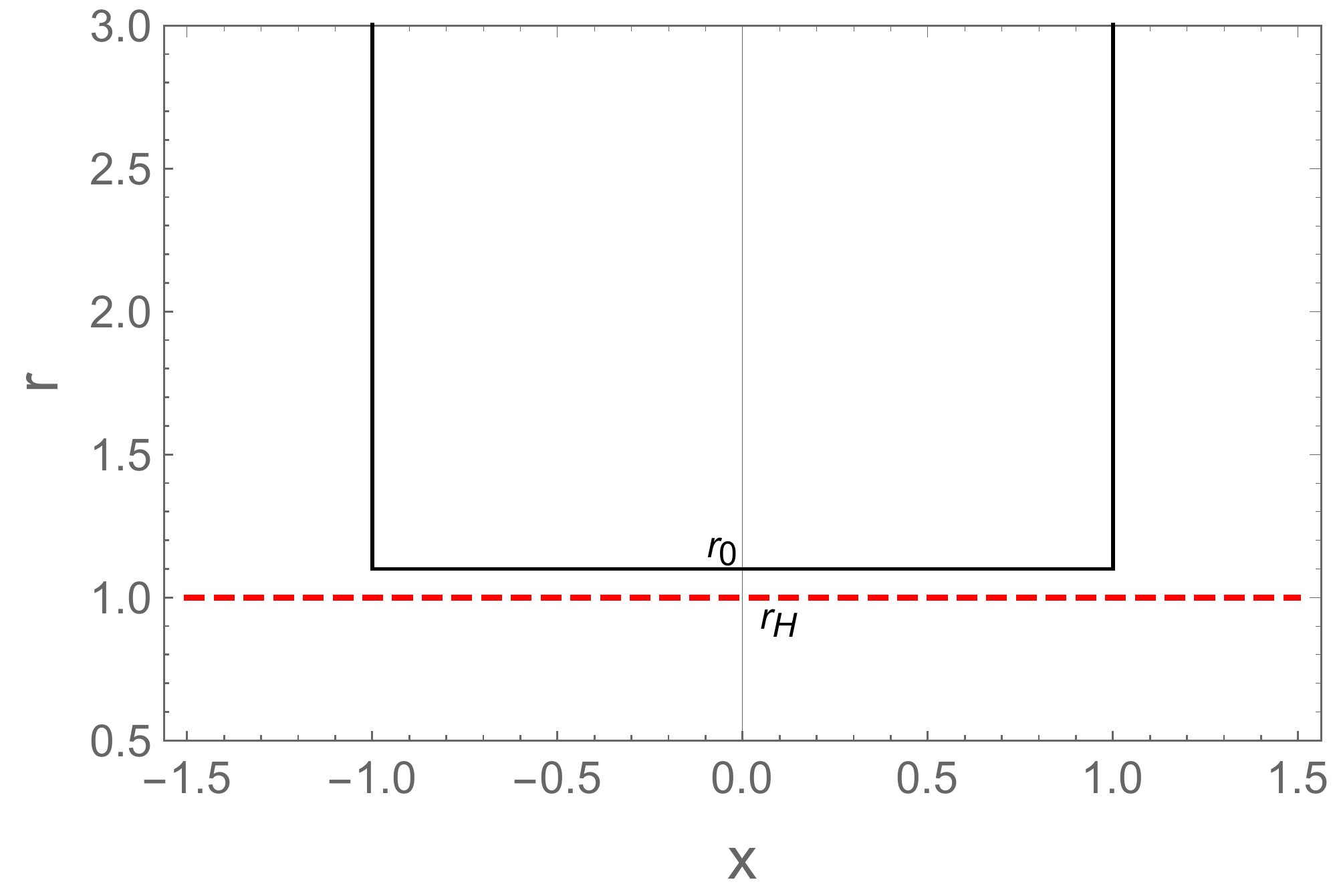}	
	\caption[\baselineskip 12 pt \small Square string in the $x$-gauge]{\baselineskip 12 pt \small Square string near the horizon, with  $r_H = 1$, $r_0 =1.1$ and $L=2$.}
	\label{Fig:2}
\end{figure}

\begin{figure}[t!]
	\centering
	\includegraphics[width=0.4 \textwidth]{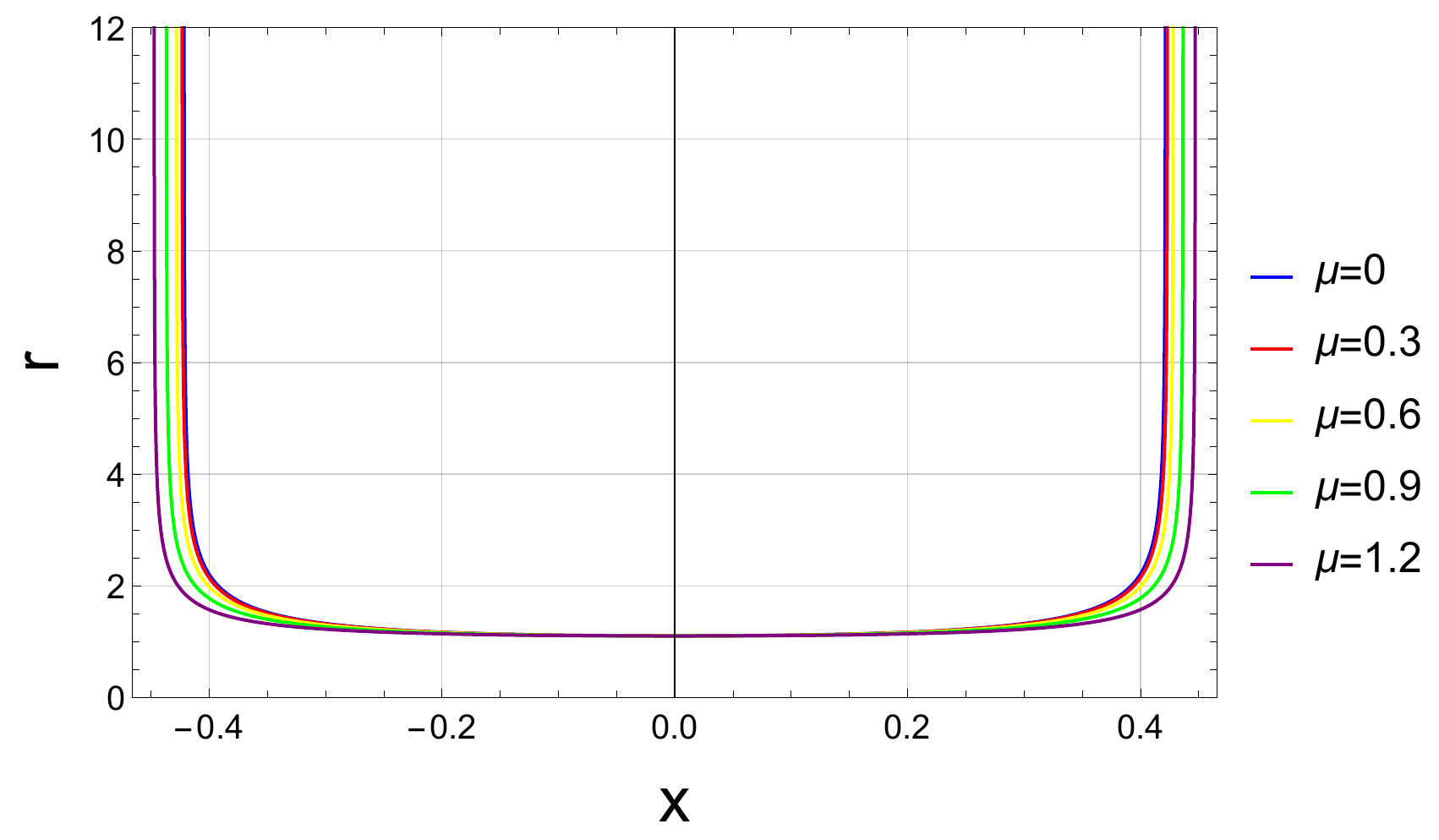}	
	\caption[Computation of a string profile]{\baselineskip 12 pt \small String profile for $r_H=1$, $r_0=1.1$ and  different values of the chemical potential $\mu$. }
	\label{Fig:3}
\end{figure}

As suggested in  \cite{Hashimoto:2018fkb}, a simple model suitable for an analytical treatment of the time-dependent perturbations is a square string in the AdS-RN background geometry \eqref{eq:21}, depicted in  Fig.~\ref{Fig:2}.
The model describes quite well  a string near the  horizon, as shown in Fig.~\ref{Fig:3} where the profile of the string  approaching the horizon is drawn. 

It is convenient to work in the $r$-gauge ($\tau = t$ and $\sigma = r$).  The embedding functions for   a string  in the $x-r$ plane are $X^{M}= \left(t,x\left( t,r \right),0,0,r \right)$, and
 the NG action reads
 \bea
&&S= - \frac{1}{2 \pi \alpha^\prime} \int \mathrm{d}t \mathrm{d}r \sqrt{1+  \acute{x}^2 \left( r^4 f \left( r \right) -\frac{1}{f \left( r \right)} \dot{r}^2 \right) }.
\nn  \\ \label{eq:31}
\eea
For a static string $X^{M}= \left(t,x\left( r \right),0,0,r \right)$ this reduces to
\begin{equation}
S=- \frac{\mathcal{T}}{2 \pi \alpha^\prime} \int \mathrm{d}r \sqrt{r^4 f \left( r \right) \acute{x}^2 + 1 }\,\,.
\label{eq:23}
\end{equation}
In the case of the square profile,  Eq.~\eqref{eq:23} is determined integrating along the three sides of the string. The result 
 can be  regularized as follows:
\begin{equation}
S^{reg} =  - \frac{\mathcal{T}}{2 \pi \alpha^\prime} \left( L r_0^2 \sqrt{ f \left( r_0 \right)} - 2 \left( r_0 - r_H \right) \right)\,\,,
\end{equation}
where $L$ still denotes the distance between the  endpoints on the boundary.
For $r_0$ near the horizon the energy  
\begin{equation}
E=-\frac{S^{reg}}{\mathcal{T}}
\label{eq:26}
\end{equation}
has a local maximum, hence upon small perturbations the string departs toward an equilibrium configuration. The  stationary point for $E$  is determined solving 
\begin{equation}
 2 L r_0 \sqrt{f\left( r_0 \right)} + \frac{r_0^2 L}{2 \sqrt{f\left( r_0 \right)}} \frac{\partial f \left( r_0 \right)}{\partial r_0} -2=0.
\label{eq:27}
\end{equation}
\noindent 
For the metric function  $f(r)$ in \eqref{eq:20},
expanding the lhs of Eq.~(\ref{eq:27})  for $r_0\to r_H$ gives:
\begin{equation}
r_{0,sol}=\frac{r_H \left( L^2 \left( 2 r_H^2 +11 \mu ^2 \right)  -8 \right)}{ 2 L^2 \left( 2 r_H ^2 +5 \mu ^2 \right) -8 }.
\label{eq:28}
\end{equation}
Moreover,  expanding for $L\to 0$ at ${\cal O} (L^2)$  gives
\begin{equation}
r_{0,sol}= r_H \left( 1+ \frac{L^2}{8} \left( 2 r_H^2 - \mu ^2 \right) \right).
\label{eq:29}
\end{equation}
\noindent 
We now consider 
a fluctuating string described by  the action in \eqref{eq:31}, and introduce
  a small time-dependent perturbation $\delta r \left( t \right)$   to the static solution, $r_0 \left( t \right) = r_{0,sol} + \delta r \left( t \right)$: indeed, 
for the square string  a perturbation makes   time-dependent the position $r_0$ of the bottom  side.
The regularized action  is given by 
\bea
S^{reg}&=&- \frac{1}{2 \pi \alpha^\prime} \int d t \Bigg\{ L \sqrt{r_0^4 f \left( r_0 \right) -\frac{1}{f \left( r_0 \right)} \dot{r_0}^2} \nn \\
&-&2 \left( r_0-r_H \right)\Bigg\}  \,\,. \label{eq:33}
\eea
The  Lagrangian 
\bea
\mathcal{L}= L \sqrt{r_0^4 f \left( r_0 \right)- \frac{1}{f \left( r_0 \right) }\dot{r}_0^2}-2 \left( r_0-r_H \right) \nn \\
\label{eq:34}
\eea
can be expanded around $r_{0,sol}$  to second order in  $\delta r( t )$:

\begin{widetext}
\begin{equation}
\begin{aligned}
\mathcal{L} \approx & -2 r_{0,sol} + 2 r_H + L \, r_{0,sol}^2 \sqrt{f \left( r_{0,sol} \right)}
+ \delta r (t) \left( -2 + 2 L r_{0,sol} \sqrt{f ( r_{0,sol})} +\frac{L r_{0,sol}^2 f^\prime ( r_{0,sol})}{2 \sqrt{f ( r_{0,sol})} } \right) \\
& +L  \, \delta r (t)^2 \Bigg(  \sqrt{f ( r_{0,sol})} + \frac{ r_{0,sol} f^\prime ( r_{0,sol})}{\sqrt{f ( r_{0,sol} )} }
- \frac{ r_{0,sol}^2 f^\prime ( r_{0,sol})^2}{8 f ( r_{0,sol})^{3/2}} + \frac{ r_{0,sol}^2 f^{\prime \prime} ( r_{0,sol})}{4 \sqrt{f ( r_{0,sol} )} }\Bigg)\\
& -\delta \dot{r} ( t )^2 \frac{L}{2 r_{0,sol}^2 f ( r_{0,sol})^{3/2} },
\end{aligned}
\label{eq:35}
\end{equation}
and the equation of motion for $\delta r \left( t \right)$ reads:
\bea
&&\delta \ddot{r} ( t ) \frac{L}{r_{0,sol}^2 f \left( r_{0,sol}\right)^{3/2}}
 + L\, \delta r ( t ) \Bigg( 2 \sqrt{f \left( r_{0,sol}\right)} + \frac{2  r_{0,sol} f^\prime \left( r_{0,sol}\right)}{\sqrt{f \left( r_{0,sol}\right)}}
- \frac{ r_{0,sol}^2 f^\prime \left( r_{0,sol}\right)^2}{4 f \left( r_{0,sol}\right)^{3/2}} + \frac{  r_{0,sol}^2 f^{\prime \prime} \left( r_{0,sol}\right)}{2 \sqrt{ f \left( r_{0,sol}\right)}} \Bigg)\nn \\
&-&2+2 \, L \, r_{0,sol} \sqrt{ f \left( r_{0,sol}\right)} + \frac{L r_{0,sol}^2 f^\prime \left( r_{0,sol}\right)}{2 \sqrt{f \left( r_{0,sol}\right)}} = 0 \,\, .
\label{eq:36}
\eea
\end{widetext}
 This equation is solved by
\begin{equation}
\delta r \left( t \right)= A \exp \left(  \lambda t \right) + B \exp \left( - \lambda t \right).
\label{eq:37}
\end{equation} 
The coefficient $\lambda$, our Lyapunov exponent,  determines the time growth  of the perturbation.  It is   given by:

\bea
&&\lambda = \frac{r_{0,sol}}{2} \Big( -8 f \left( r_{0,sol} \right) ^2 +r_{0,sol} ^2 f^\prime \left( r_{0,sol} \right) ^2 \nn \\ 
&-& 2 r_{0,sol} f \left( r_{0,sol} \right) \big( 4 f^\prime \left( r_{0,sol}\right)  
+ r_{0,sol} f^{\prime \prime} \left( r_{0,sol} \right) \big) \Big) ^{1/2}. \nn \\
\label{eq:38}
\eea
Expanding $f \left( r_{0,sol} \right)$, $f^\prime \left( r_{0,sol} \right)$ and $f^{\prime \prime} \left( r_{0,sol} \right)$
at  second order in $L$ we have:

\begin{equation}
\lambda = 2 r_H \left( 1- \frac{\mu^2}{2 r_H^2} \right) \left( 1-\frac{L^2}{4} \left( 2 r_H^2 - \mu^2 \right) \right) \,\,.
\label{eq:40}
\end{equation}
Using Eq.~\eqref{eq:thBH}   we find

\begin{equation}
\lambda = 2 \pi T_H \left( 1-\frac{L^2}{2} \pi T_H r_H \right).
\label{eq:41}
\end{equation}
 The exponent $\lambda$ saturates the  bound \eqref{eq:1} at the lowest order in $L$. 
 The ${\cal O}(L^2)$   correction is negative:   Eq.\eqref{eq:41} can be written as
 
\bea
\lambda = 2 \pi T_H \left( 1-\frac{L^2}{4} \pi^2 T_H ^2 \left( 1+ \sqrt{1+\frac{2 \mu^2}{\pi^2 T_H^2}}\right) \right) , \nn \\
\label{eq:41bis}
\eea
hence  the coefficient of the $L^2$  correction increases with $\mu$.

 \section{Perturbed string}
To study the onset of chaos in a more realistic configuration, we perturb the static solution of a string near the black-hole horizon by a small time-dependent effect. 

\begin{figure}[t!]
	\centering
%	\vspace*{1.5cm}
	\includegraphics[width=0.35 \textwidth]{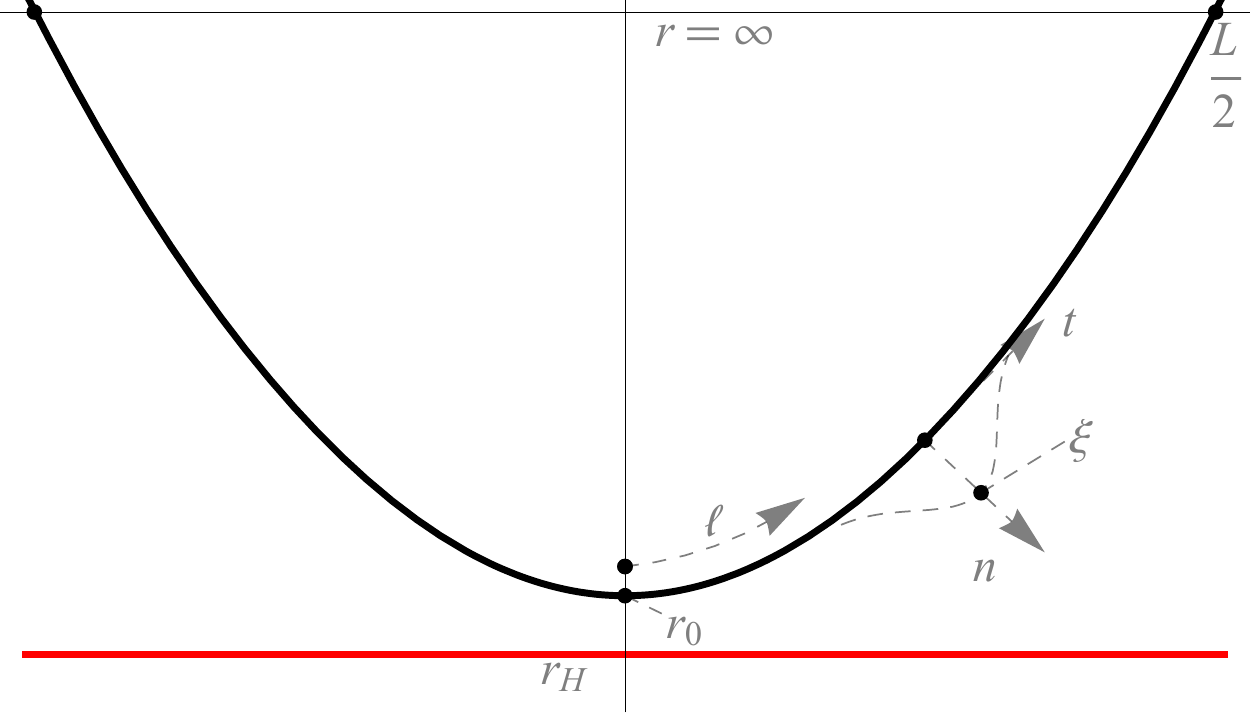}
	\caption[Perturbed string]{\baselineskip 12 pt \small Perturbation along the direction orthogonal to the string in each point with coordinate $l$.}
	\label{Fig:6}
\end{figure}

There are  different ways to introduce a small time-dependent perturbation. We follow  \cite{Hashimoto:2018fkb}, and perturb the string along the orthogonal direction at each point with coordinate $l$ in the $r-x$ plane, as in Fig.~\ref{Fig:6}.
For the unit vector $n^M= (0,n^x,0,0,n^r)$ orthogonal to $t^M$ we have:

\bea
g_{rr} ( r ) \left({n}{^{r}}\right)^2 +g_{xx} ( r) \left({n}{^{x}}\right)^2 &=& 1
\label{eq:49} \\
\acute{r} \left( l \right)  g_{rr} \left( r \right)  \, n^r + \acute{x} \left( l \right)  g_{xx}\left( r \right)  \, n^x&=&0.
\label{eq:48}
\eea
The solution for the components $n^{x}$ and $n^{r}$  is

\begin{equation}
{n^x} ( l)=\sqrt{\frac{g_{rr}}{g_{xx}}} \; \acute{r}( l ) \quad \text{,} \quad {n}{^{r}} ( l )=-\sqrt{\frac{g_{xx}}{g_{rr}}} \; \acute{x}( l )
\label{eq:50}
\end{equation}
\noindent 
for an outward perturbation, as  in Fig.~\ref{Fig:6}. 
Introducing a time-dependent perturbation $\xi \left( t,l \right)$ along $n$ one has:

\bea
r \left( t,l \right) &=& r_{BG} \left( l \right) + \xi \left( t,l \right) n^{r} \left( l \right), \nn \\
x\left( t,l \right) &= & x_{BG} \left( l \right) + \xi \left( t,l \right) n^{x} \left( l \right),
\label{eq:51} \\ \nn
\eea
with $r_{BG} \left( l \right)$ and $x_{BG} \left( l \right)$  the static solutions obtained integrating Eqs.~(\ref{eq:46}) and (\ref{eq:47}).

To describe the dynamics of the perturbation (assuming it is small), we expand the metric function around the static solution $r_{BG}(l)$  to the third order in  $\xi \left( t,l \right)$.

\begin{figure}[h]
\centering
\includegraphics[width=0.35 \textwidth]{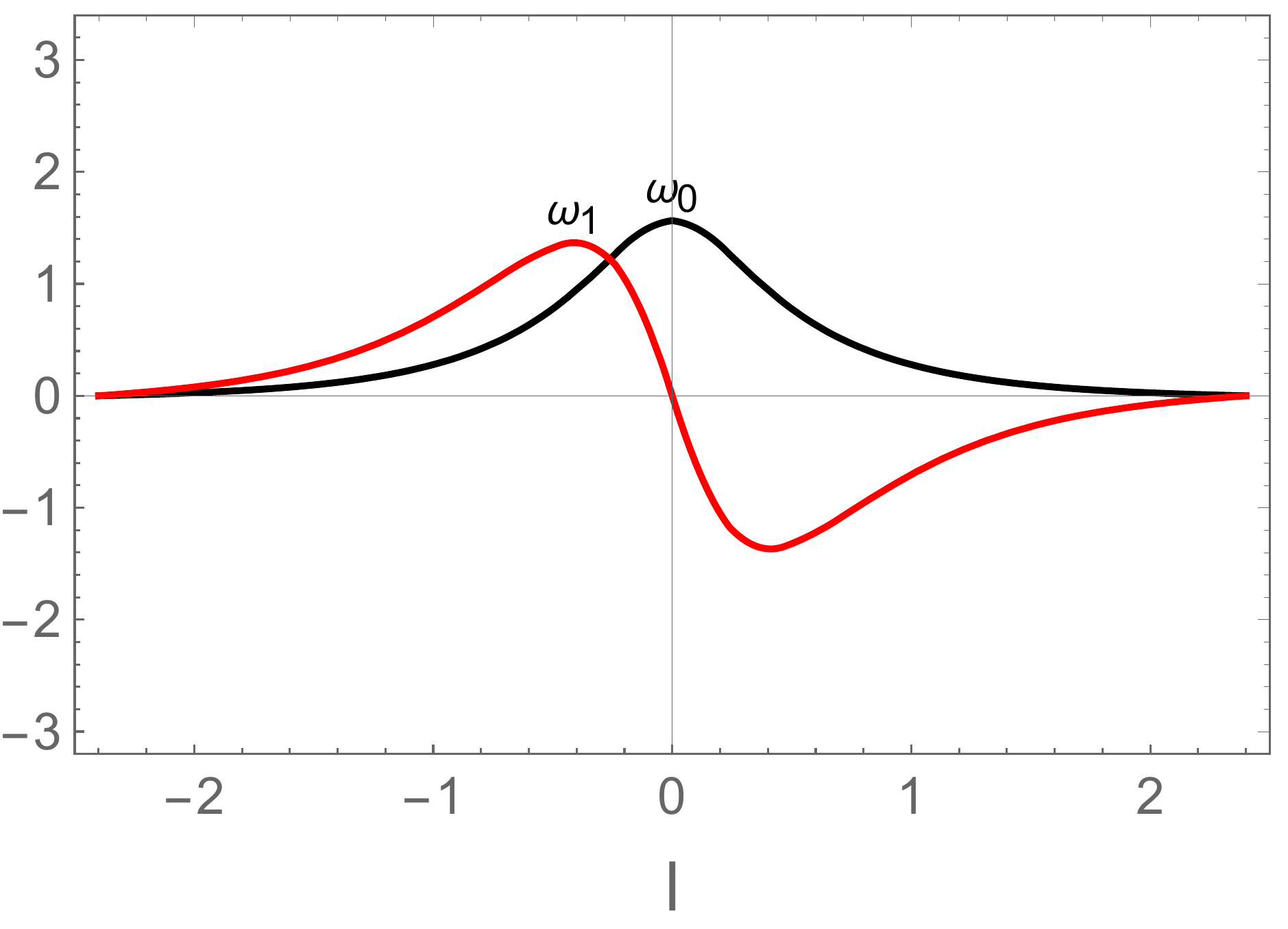}
\caption{ \baselineskip 12 pt \small Eigenfunctions $e_0 \left( l \right)$ (black line) and $e_1 \left( l \right)$ (red line) of  Eq.~(\ref{eq:58}) for  $r_0=1.172$ and $\mu=0.6$.}\label{Fig:eigen}
\end{figure}

To the third order in $\xi$ the NG action involves a quadratic and a cubic term.
The  quadratic term  has the form:

\bea
&&S^{\left( 2 \right)} = \nn \\
&&\frac{1}{2 \pi \alpha^\prime}\int \mathrm{d}t \int_{-\infty}^{\infty} \mathrm{d}l \left( C_{tt}  \dot{\xi}^2 + C_{ll}  \acute{\xi}^2  + C_{00} \xi^2\right),
\nn \\
\label{eq:54} 
\eea
\noindent
with $C_{tt}$, $C_{ll}$ and $C_{00}$ depending on $l$. 
For the metric in Eq.~(\ref{eq:21}) with a generic metric function $f(r)$ the coefficients $C_{tt}$, $C_{ll}$ and $C_{00}$ read:

\begin{widetext}
\begin{equation}
\begin{aligned}
C_{tt} \left( l \right)&=\frac{1}{2 r_{BG} \sqrt{f \left( r_{BG} \right) }},\\
C_{ll} \left( l \right)&=  - \frac{1}{4 C_{tt} \left( l \right) } , \\
C_{00} \left( l \right) &= \frac{1}{4 r_{BG}^3 f ( r_{BG} )^{
 3/2}}\Bigg\{ \Big( -2 r_{BG}^4 f ( r_{BG} ) ^2 \big( 2 f \left( r_{BG} \right) 
+ r_{BG} f^\prime \left( r_{BG} \right) \Big) \\
&+ r_0^4 f \left( r_0 \right) \Big( 4 f \left( r_{BG} \right)^2 + r_{BG}^2 f^\prime \left( r_{BG} \right) ^2
+ r_{BG} f \left( r_{BG} \right) \left( f^\prime \left( r_{BG} \right) - r_{BG} f^{\prime \prime} \left( r_{BG} \right) \right) \Big) \Bigg\}.
\end{aligned}
\label{eq:55}
\end{equation}
The coefficients depend on $l$ through $r_{BG} \left( l \right)$. Their expressions for the AdS-RN metric are:

\bea
C_{tt} \left( l \right)&=&\frac{r_{BG}^2}{2 \sqrt{\left( r_{BG}^2 -r_H^2 \right) \left( r_{BG}^4  + r_{BG}^2 r_H^2 - r_H^2 \mu^2 \right)}}, \nn\\
C_{ll} \left( l \right)&=& - \frac{1}{4 C_{tt} \left( l \right)}, \nn\\
C_{00} \left( l \right) &=&  \bigg( r_0^6 r_{BG}^2 \big( r_{BG}^{12} - 10 r_{BG}^6 r_H^4 \mu^2 - 2 r_H^8 \mu^4 + 4 r_{BG}^8 r_H^2 ( r_H^2 
+ \mu^2 ) + 4 r_{BG}^2 r_H^6 \mu^2 ( r_H^2 + \mu^2 ) - r_{BG}^4 r_H^4 ( r_H^2 + \mu^2 )^2 \big) \nn\\
&+& r_{BG}^2 r_H^4 \mu^2 \big( r_{BG}^{12} - 10 r_{BG}^6 r_H^4 \mu^2 - 2 r_H^8 \mu^4 + 4 r_{BG}^8 r_H^2 ( r_H^2 + \mu^2 )+ 4 r_{BG}^2 r_H^6 \mu^2 ( r_H^2 + \mu^2 ) \nn\\
& - &r_{BG}^4 r_H^4 ( r_H^2 + \mu^2 )^2 \big) 
- r_0^2 \big( r_{BG}^{18} - 3 r_{BG}^6 r_H^8 \mu^4 - 2 r_H^{12} \mu^6 - 6 r_{BG}^8 r_H^6 \mu^2 ( r_H^2 + \mu^2 ) \nn \\
&+ &3 r_{BG}^2 r_H^{10} \mu^4 ( r_H^2 + \mu^2 ) 
+ 3 r_{BG}^{10} r_H^4 ( r_H^2 + \mu^2)^2 ) \big) \bigg)  \frac{1}
{  r_0^2 \, r_{BG}^{8} \bigg( ( r_{BG}^2 - r_H^2 ) ( r_{BG}^4 + r_{BG}^2 r_H^2 - r_H^2 \mu^2 ) \bigg) ^{3/2}  }. 
\label{eq:56}
\eea
\end{widetext}
The  equation of motion from \eqref{eq:54}  is

\begin{equation}
C_{tt} \, \ddot{\xi} + \partial_l \left( C_{ll} \acute{\xi} \right) - C_{00} \, \xi = 0.
\label{eq:57}
\end{equation}
For $\xi \left( t,l \right) = \xi \left(l \right) e^{i \omega t}$  this corresponds to

\be
\partial_l \left( C_{ll} \, \acute{\xi} \right) - C_{00} \, \xi = \omega ^2 C_{tt}  \, \xi \,\, ,
\label{eq:58}
\ee
a Sturm-Liouville equation with weight function $W(l)=-C_{tt}(l)$.
We solve Eq.~\eqref{eq:58} for different values of $r_0$ and $\mu$, imposing the boundary conditions $ \xi \left( l \right) \xrightarrow{ l \rightarrow \pm \infty } 0$. The  two lowest eigenvalues $\omega_0^2$ and $\omega_1^2$, varying $r_0$ and $\mu$,  are collected in Table~\ref{Tab:1}, and  in one case the eigenfunctions $e_0 \left(l \right)$ and $e_1 \left(l \right)$ are depicted in Fig.~\ref{Fig:eigen}.

\begin{table}[b!]
\centering
\begin{tabular}{c c c c | c c c c}
\\ \\
\toprule
&$r_0=1.1$&       			 	&		&&$r_0=1.172$&       				& 		 \\
&$\mu$       & $\omega_0^2$		& $\omega_1^2$  & &$\mu$       & $\omega_0^2$	& $\omega_1^2$  \\
&0		   & -1.370				& 7.638    &&0			& -0.064				& 10.458	\\
&0.3		   & -1.235				& 7.418	&&0.3		& -0.005				& 10.239	\\
&0.6		   & -0.870				& 6.748	&&0.6		& 0.148				& 9.574	\\
&0.9	           & -0.388				& 5.605	&&0.9		& 0.324				& 8.428	\\
&1.2		  & 0.006				& 3.938	&&1.2		& 0.397				& 6.735	\\
\toprule
&$r_0=1.18$&      				&		&&$r_0=5$&       		                         & 		 \\
&$\mu$       & $\omega_0^2$		& $\omega_1^2$  & &${\mu}$& $\omega_0^2$		& $\omega_1^2$  \\
&0			& 0.071			& 10.754	\,\,\,&&0	        & 81.726		& 275.477	 \\
&0.3			& 0.124			& 10.537	      &&0.3		& 81.706		& 275.458	 \\
&0.6			& 0.258			& 9.874	      &&0.6		& 81.648		& 275.400	 \\
&0.9			& 0.406			& 8.733	      &&0.9		& 81.551		& 275.303	 \\
&1.2			& 0.449			& 7.046	      &&1.2		& 81.415		& 275.168	 \\
\toprule
\end{tabular}
\caption[Values of the first two eigenvalues]{\baselineskip 12 pt \small Eigenvalues $\omega_0^2$ and $\omega_1^2$ of Eq.~\eqref{eq:58} changing the values of $r_0$ and $\mu$.}\label{Tab:1}
\end{table}

\begin{figure}[t!]
\centering
	\includegraphics[width=0.35 \textwidth]{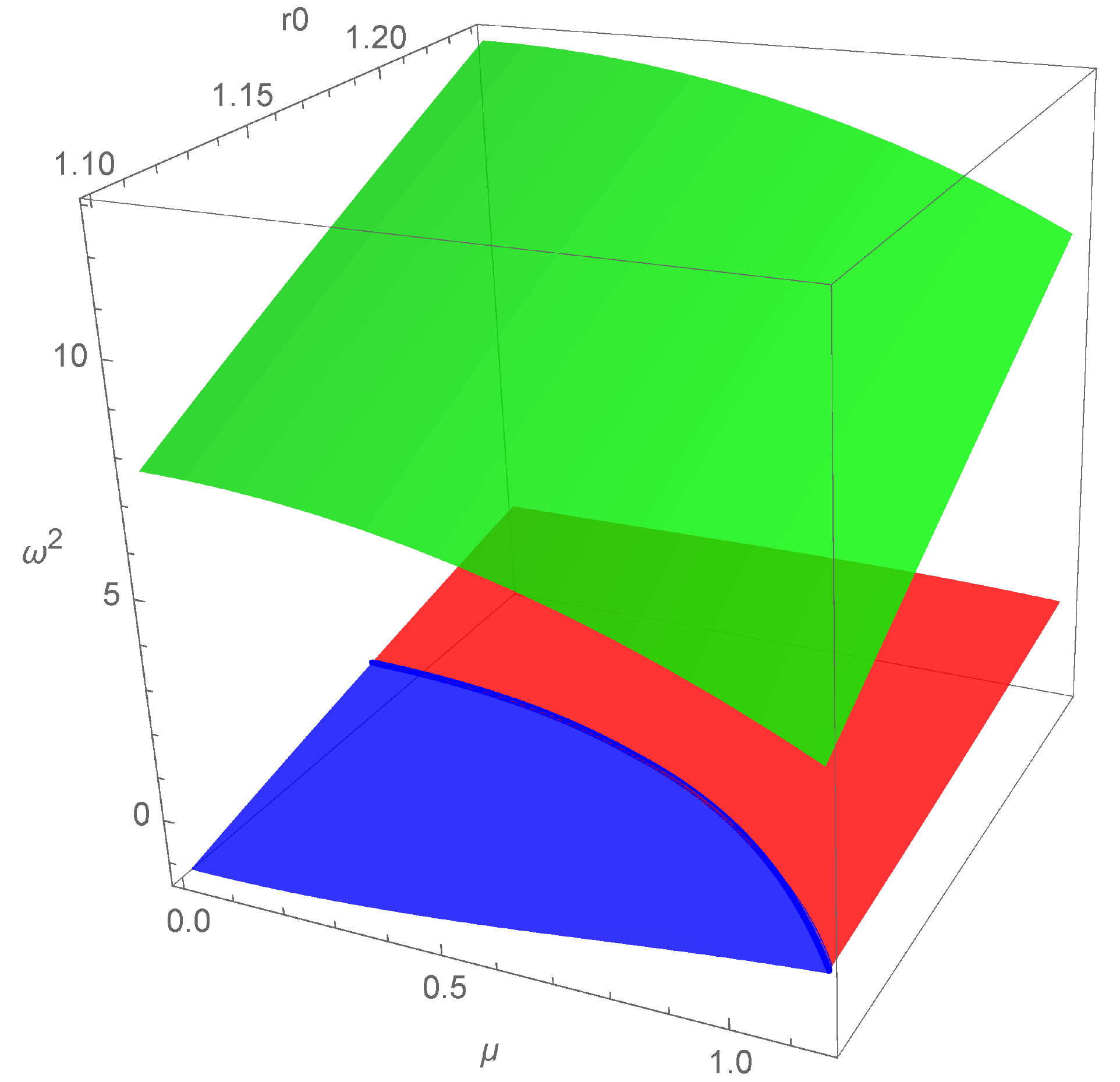}
	\caption[Eigenvalues vs $\mu$ and $r_0$]{\baselineskip 12 pt \small Eigenvalues $\omega_0^2$ and $\omega_1^2$ vs $r_0$ and $\mu$. The green surface corresponds to  $\omega_1^2$, the red and blue surface to $\omega_0^2$. The dark blue line demarcates the  (blue) region of negative $\omega_0^2$ from the (red) region of positive $\omega_0^2$.}
\label{Fig:9}
\end{figure}

There are negative values of $\omega_0^2$, corresponding to an unstable sector. For $\mu = 0$ the system is stabilized as  $r_0$  increases, with the tip of the string departing from the BH horizon:  $\omega_0^2$ becomes positive for $r_0\geq 1.177$. Fixing $r_0 = 1.1$, the lowest lying state is stabilized  increasing the chemical potential $\mu$,  and $\omega_0^2$  is positive for $\mu\geq 1.2$.  The dependence of $\omega_0^2$ and $\omega_1^2$ on $r_0$ and $\mu$ is shown in Fig.~\ref{Fig:9}, together with  the line demarcating the regions of  negative and positive values of $\omega_0^2$.

The  perturbation can be expanded in  terms of the first two eigenfunctions  $e_0$ and $e_1$,

\begin{equation}
\xi \left( t,l \right) = c_0 \left( t \right) e_0 \left( l \right) + c_1 \left( t \right) e_1 \left( l \right),
\label{eq:60}
\end{equation} 
\noindent 
with the time dependence dictated by  the coefficient functions $c_0( t )$ and $c_1 ( t )$. 
Up to a surface term, the  cubic action has the  expression:
\begin{widetext}
\be
S^{\left( 3 \right)} =  \frac{1}{2 \pi \alpha^\prime}\int \mathrm{d}t \int_{-\infty}^{\infty} \mathrm{d}l \bigg\{ D_0 \,   \xi^3 
+ D_1 \, \xi \acute{\xi}^2 + D_2 \, \xi \dot{\xi}^2  \bigg \} \,\, ,\label{eq:62}
\ee
with $D_{0,1,2}$ functions of $l$. This  reads,  expanding  the perturbation $\xi(t,l)$ as in  Eq.~(\ref{eq:60}):

\bea
S^{\left( 3 \right)}&=&
 \frac{1}{2 \pi \alpha^\prime}\int \mathrm{d}t \int_{-\infty}^{\infty} \mathrm{d}l \Bigg\{ \Big( D_0  \,  e_0^3 + D_1 \, e_0 \acute{e}_0^2 \Big) c_0^3 \left( t \right) \nn \\
 &+& \Big( 3 D_0   \, e_0 e_1^2 + D_1  \left( 2 \acute{e}_0 e_1 \acute{e}_1 + e_0 \acute{e}_1^2 \right)\Big) c_0 \left( t \right) c_1^2 \left( t \right)
 + D_2  \Big( e_0 e_1^2 c_0 \dot{c}_1^2 +e_0^3 e_1^2 c_0 \dot{c}_0^2+ 2 e_0 e_1^2 \dot{c}_0 c_1 \dot{c}_1    \Big) \Bigg\}. 
\label{eq:63}
\eea

\noindent 
Upon integration on $l$,  an action for $c_0(t)$ and $c_1(t)$ is obtained  summing $S^{(2)}$ and $S^{(3)}$:
\be
S^{(2)}+S^{(3)}=\frac{1}{2 \pi \alpha^\prime}\int \mathrm{d}t \Bigg[ \sum_{n=0,1}\left(\dot{c}_n^2-\omega_n^2 c_n^2  \right) + K_1 c_0^3 
+ K_2 c_0 c_1^2  + K_3 c_0 \dot{c}_0^2 + K_4 c_0 \dot{c}_1^2 + K_5 \dot{c}_0 c_1 \dot{c}_1\Bigg].
\label{eq:64}
\ee
\end{widetext}

\noindent 
The  coefficients $K_{1, \dots 5}$ depend on   $r_0$ and $\mu$, and are collected in Table~\ref{Tab:3} choosing a set of values for  such quantities.

\begin{table}[b!]
\centering
\begin{tabular}{ccccccc}
\\
\toprule
$r_0=1.1$\quad &{$\mu$}		& {$K_1$}		& {$K_2$}		&{$K_3$}		&{$K_4$}		&{$K_5$}  \\
&0			& 11.36			& 21.72			& 10.58			& 3.37			& 6.73	      \\
&0.6			& 7.22			& 16.76			& 9.98			& 3.44			& 6.88	      \\
&1.2			& 0.81			& 5.84			& 8.29			& 3.64			& 7.28	      \\
\toprule
$r_0=1.172$ \quad&{$\mu$}		& {$K_1$}		& {$K_2$}		&{$K_3$}		&{$K_4$}		&{$K_5$}  \\
&0			& 7.63			& 20.61			& 8.17			& 2.69			& 5.39	      \\
&0.6			& 5.13			& 17.30			& 8.04			& 2.81			& 5.62	      \\
&1.2			& 0.86			& 9.30			& 7.81			& 3.22			& 6.44	      \\
\toprule
$r_0=1.18$\quad &{$\mu$}		& {$K_1$}		& {$K_2$}		&{$K_3$}		&{$K_4$}		&{$K_5$}  \\
&0			& 7.36			& 20.64			& 8.00			& 2.65			& 5.29	      \\
&0.6			& 4.97			& 17.45			& 7.90			& 2.76			& 5.53	      \\
&1.2			& 0.88			& 9.69			& 7.76			& 3.18			& 6.36	      \\
\toprule
$r_0=5$\quad &{$\mu$}		& {$K_1$}		& {$K_2$}		&{$K_3$}		&{$K_4$}		&{$K_5$}  \\
&0			& -15.01      	& 560.52		& 7.44		    & 2.84			& 5.67	      \\
&0.6			& -14.88		& 560.57		& 7.44			& 2.84		& 5.67	      \\
&1.2			& -14.49		& 560.73		& 7.46			& 2.84		& 5.69	      \\
\toprule
\end{tabular}
\caption[ $K_{1,\dots,5}$ coefficients of the third order action]{\baselineskip 12 pt \small $K$ coefficients in Eq.~(\ref{eq:64}) changing the  values of $r_0$ and  $\mu$.}\label{Tab:3}
\end{table}

As one can  numerically test, in cases corresponding  to  negative values of $\omega_0^2$ the action describes the motion of $c_0$ and $c_1$ in a trap,
and in some regions within the  potential the kinetic term is negative. As suggested in \cite{Hashimoto:2018fkb},  it is useful to replace $c_{0,1}\to \tilde c_{0,1}$ in the action, with
%
%\begin{equation}
%c_0=\tilde{c}_0 + \alpha_1 \tilde{c}_0^2 + \alpha_2 \tilde{c}_1^2, \qquad c_1 = \tilde{c}_1 + \alpha_3 \tilde{c}_0 \tilde{c}_1  
%\label{eq:65}
%\end{equation}
%
$c_0=\tilde{c}_0 + \alpha_1 \tilde{c}_0^2 + \alpha_2 \tilde{c}_1^2$ and $c_1 = \tilde{c}_1 + \alpha_3 \tilde{c}_0 \tilde{c}_1$,
neglecting  $\mathcal{O} \left( \tilde{c}_i^4 \right)$ terms, setting the constants $\alpha_i$  to ensure the positivity of the kinetic term. We set the constants $\alpha_1=-2$, $\alpha_2 = -0.5$ and $\alpha_3 = -1$, slightly different from  \cite{Hashimoto:2018fkb}. The replacement stretches the potential  stabilizing the time evolution: the dynamics is not affected, and a chaotic behaviour  shows up also in the transformed system.

\begin{figure*}[t!]
\centering
\makebox[\linewidth][c]{{
	{\includegraphics[width=0.35 \textwidth]{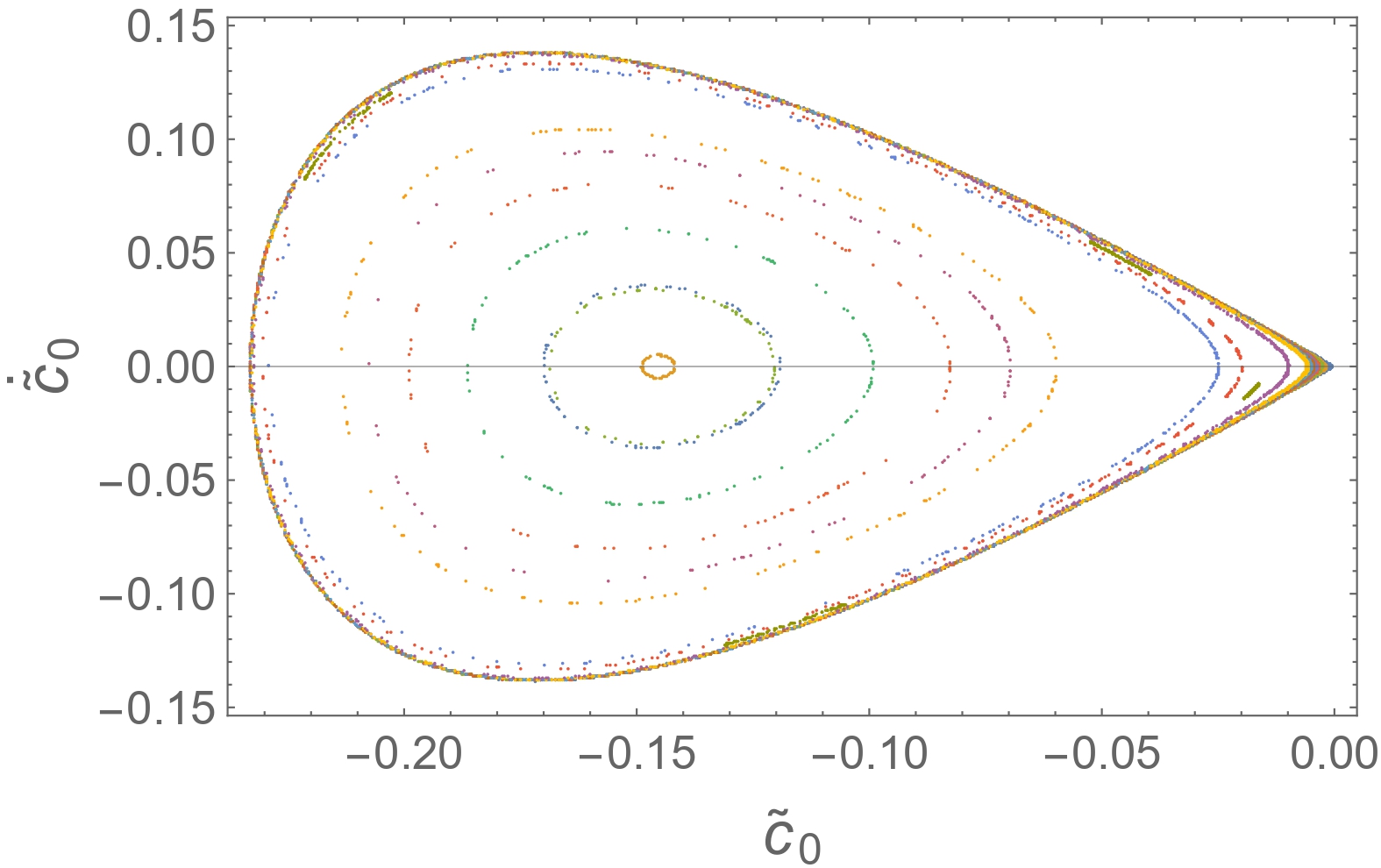} \quad
	 \includegraphics[width=0.37 \textwidth]{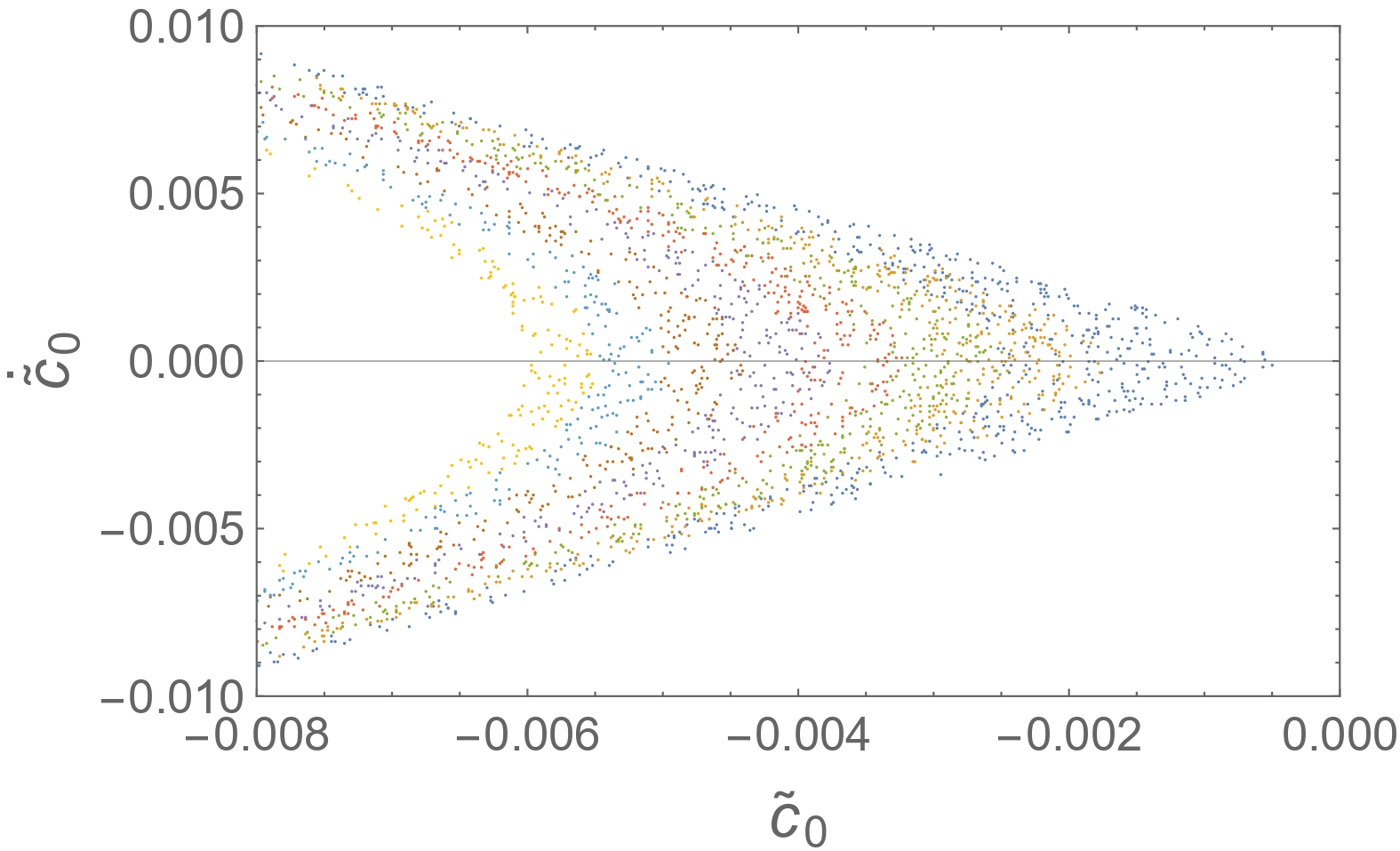}}}} \quad
\makebox[\linewidth][c]{{
	{\includegraphics[width=0.35 \textwidth]{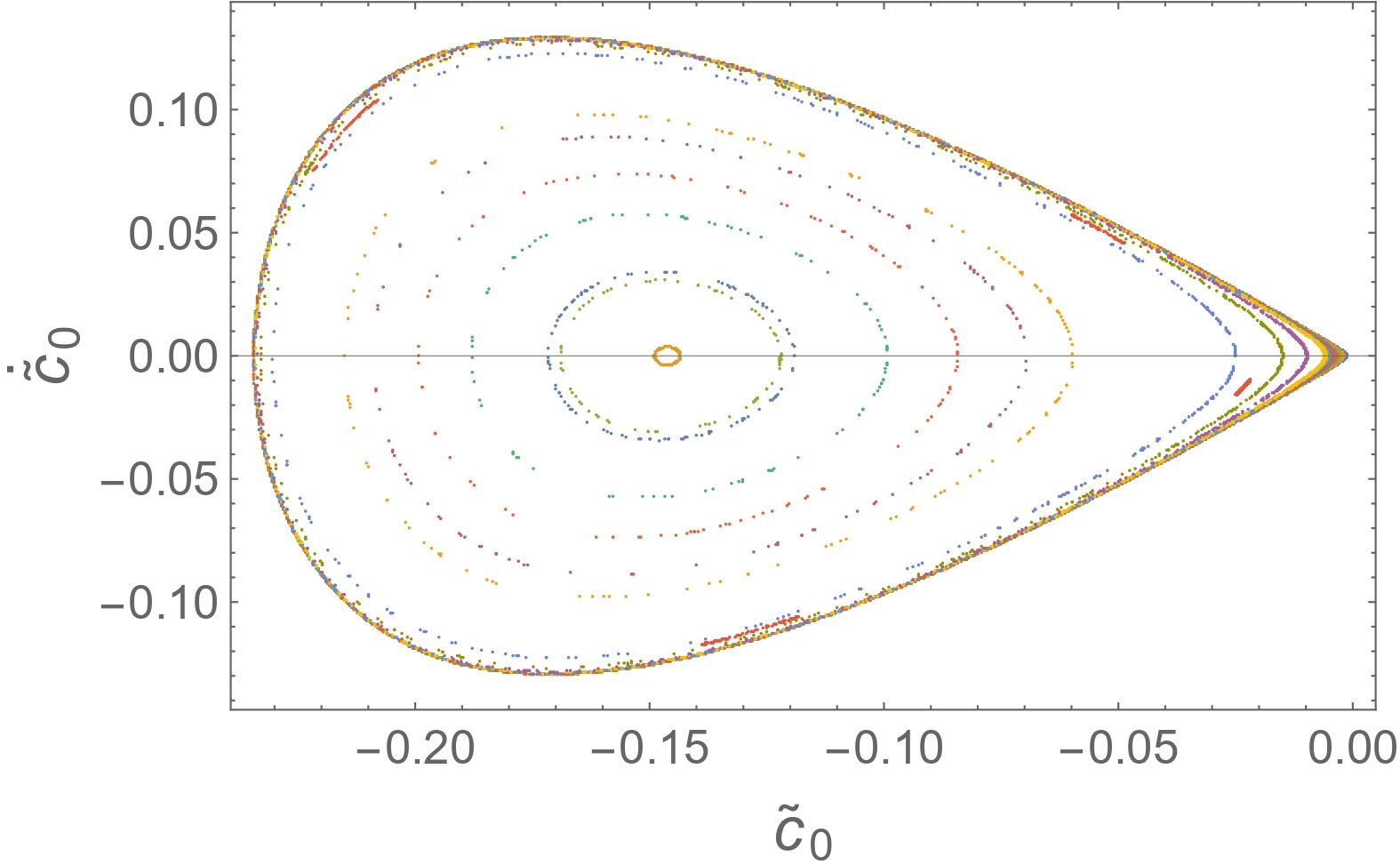} \quad
	 \includegraphics[width=0.37 \textwidth]{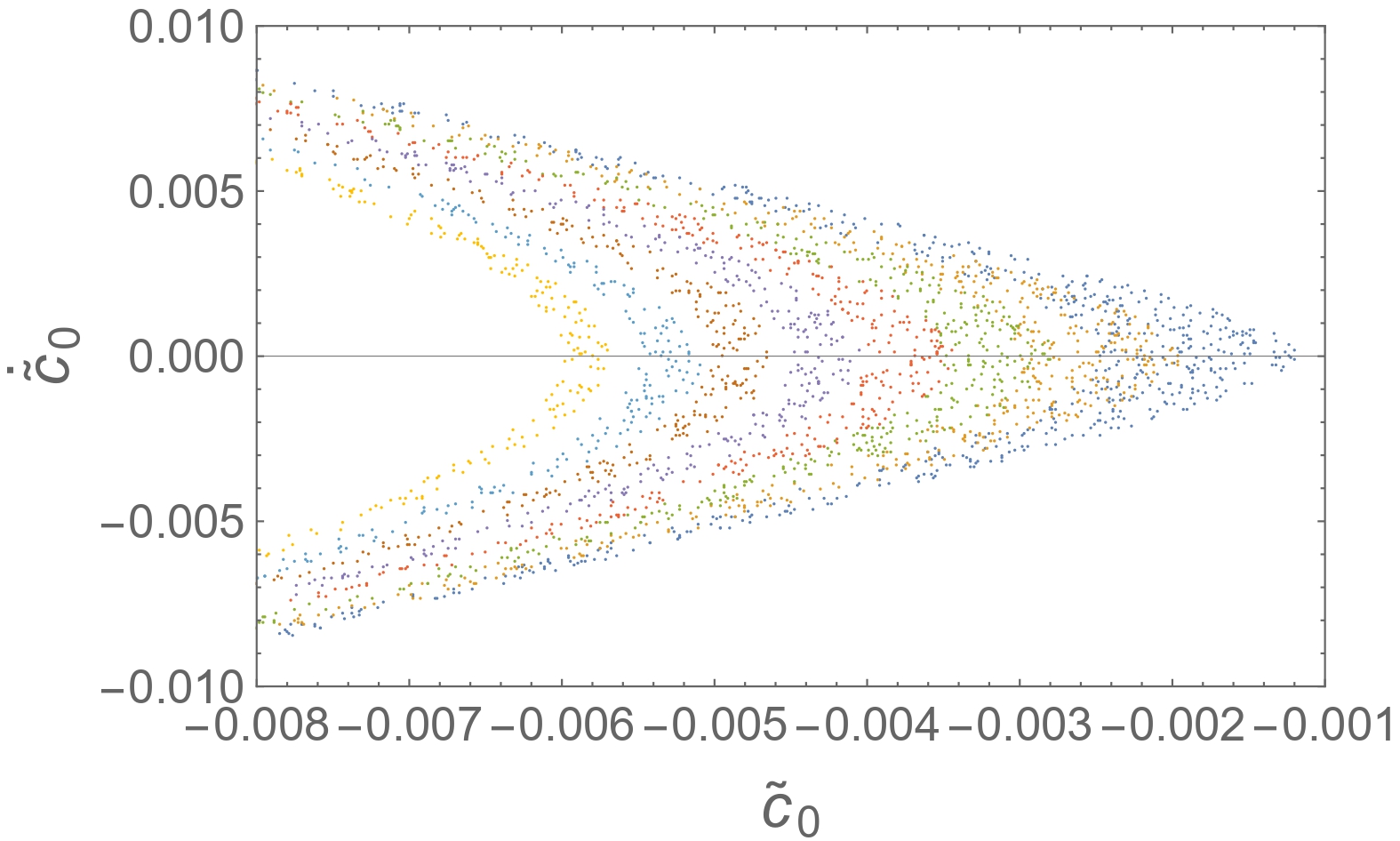}}}} \quad
\makebox[\linewidth][c]{{
	{\includegraphics[width=0.35 \textwidth]{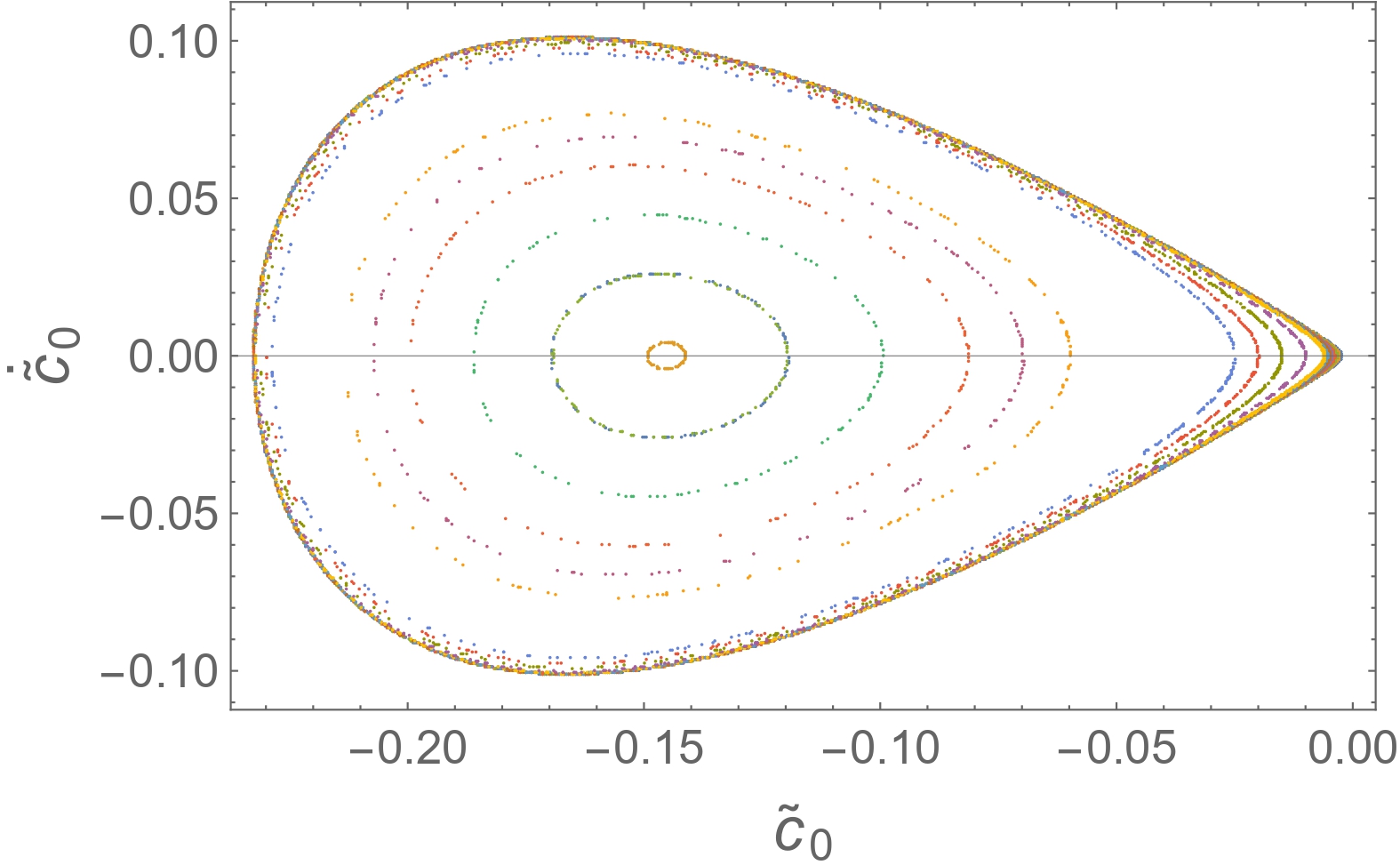} \quad
	 \includegraphics[width=0.37 \textwidth]{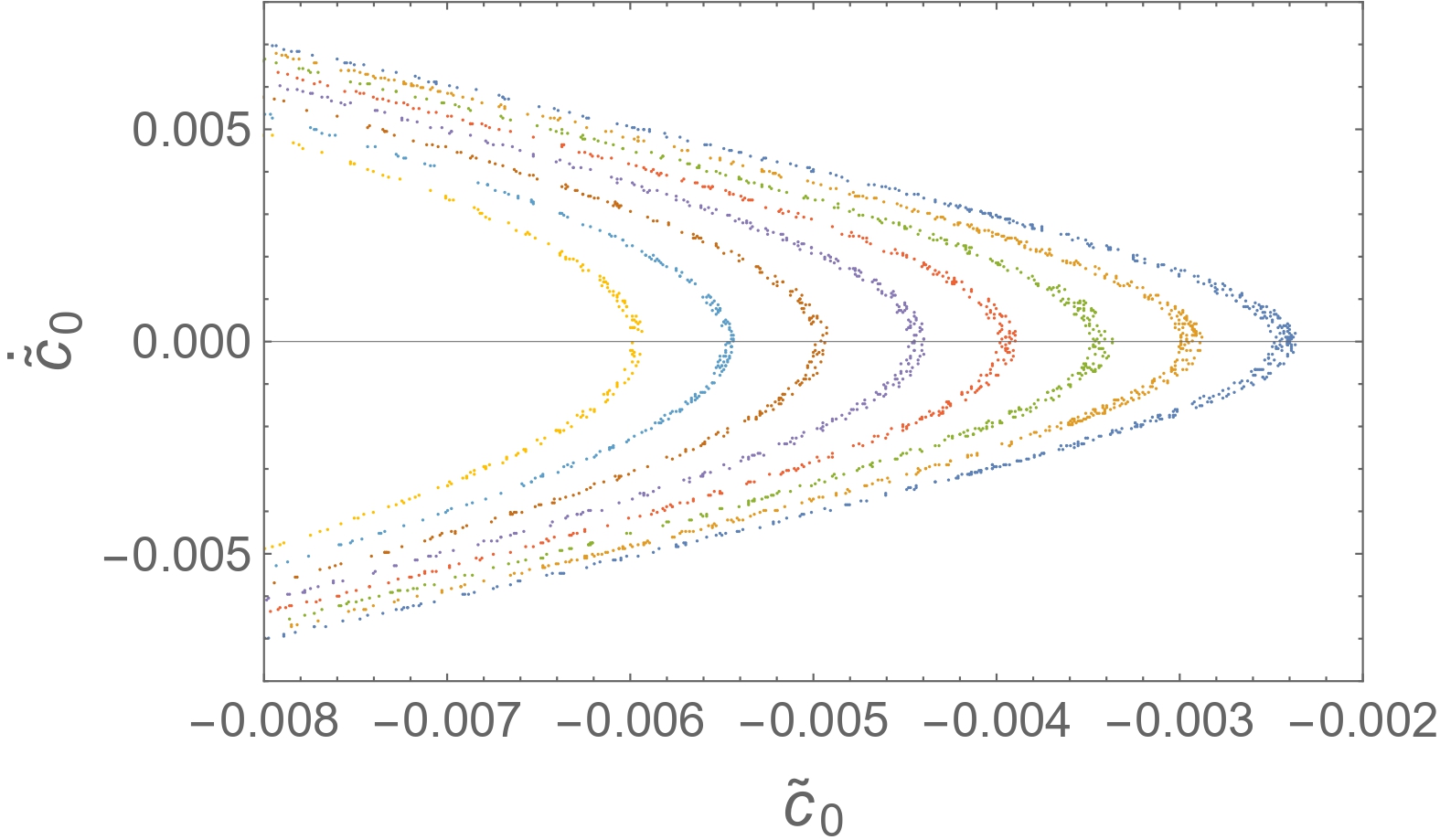}}}} \quad
\makebox[\linewidth][c]{{
	{\includegraphics[width=0.35 \textwidth]{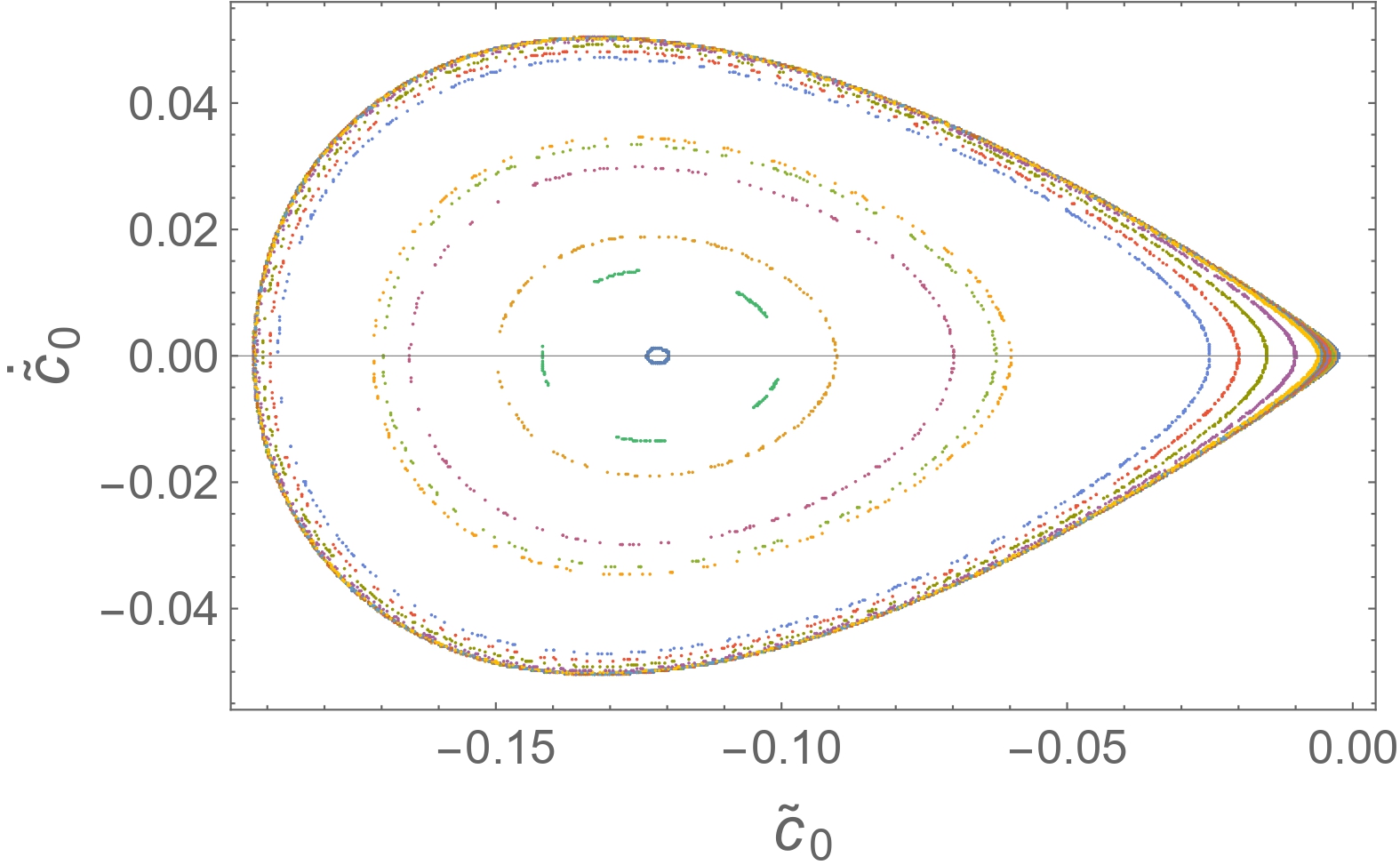} \quad
	 \includegraphics[width=0.37 \textwidth]{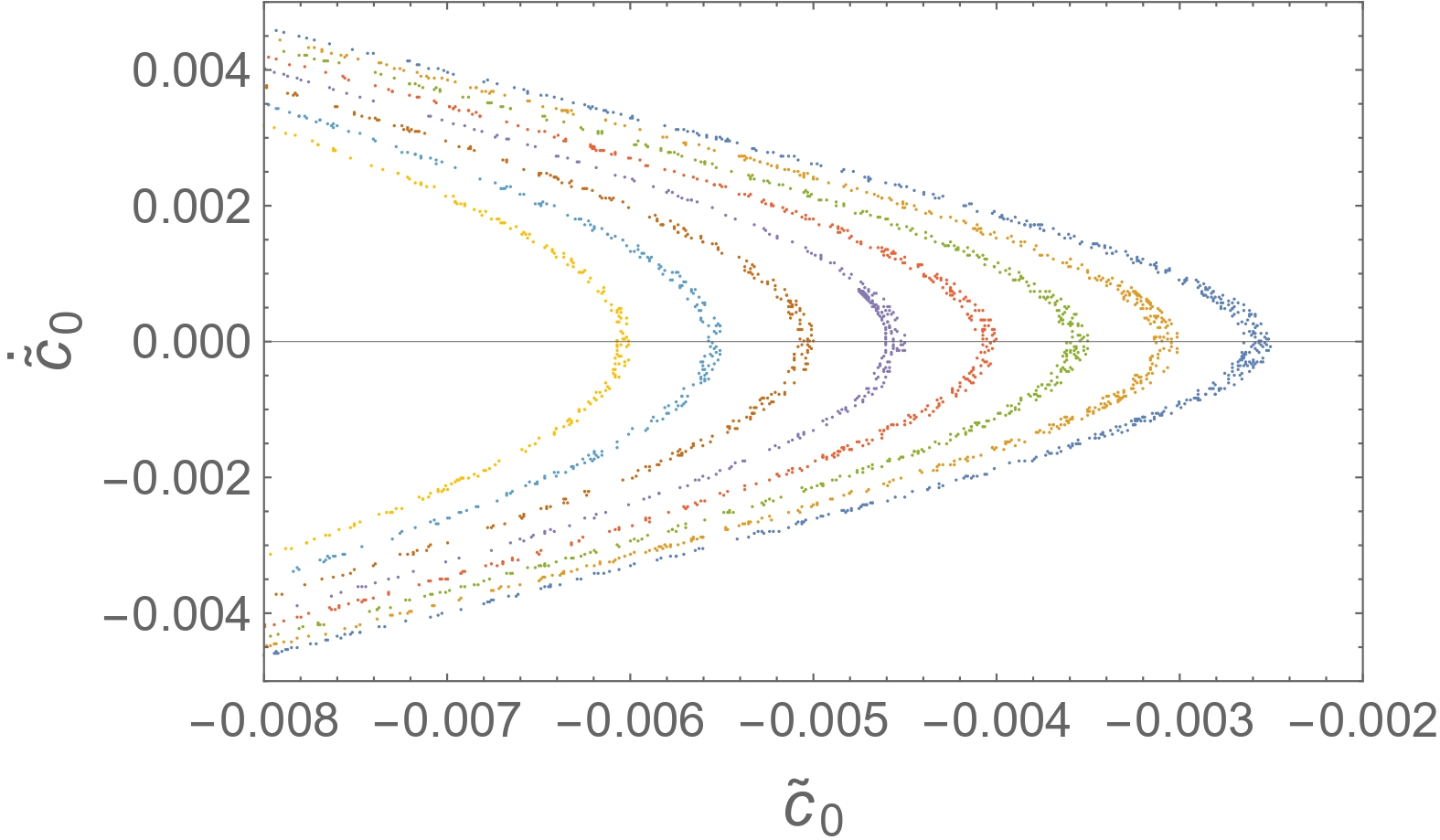}}}}
\caption[Poincar\'e plots for $r_0 = 1.1$ and different value of $\mu$]{\baselineskip 12 pt \small Poincar\'e sections for a time-dependent perturbed string, obtained changing the initial conditions, with $r_0=1.1$ and increasing the chemical potential:  $\mu=0$ (top row), $\mu=0.03$ (second row), $\mu=0.06$ (third row) and $\mu=0.09$ (bottom row), for $\tilde{c}_1 =0$ and $\dot{\tilde{c}}_1 \geq 0$. The plots in the right column enlarge the corresponding ones in the left column in the range of small  $\tilde{c}_0$, $\dot{\tilde{c}}_0$.}\label{Fig:12}
\end{figure*}

 To gain information on  chaos  we adopt a  procedure analogous to the one in  Sec.~\ref{square}: we start considering  a static solution and perturb it with a small time-dependent fluctuation. However, in this case  an analytic computation as in the simplified case in  Sec.~\ref{square} cannot be used.
Onset of chaos can be investigated constructing  Poincar\'e sections numerically. We show the sections defined by $\tilde{c}_1 \left( t \right) = 0$ and $\dot{\tilde{c}}_1 \left( t \right) > 0$ for bounded orbits within the  trap. In the case $r_H=1$, $r_0 = 1.1$ and increasing $\mu$ such sections are collected  in Fig.~\ref{Fig:12}. For  $\tilde{c}_0$ near zero the orbits are scattered points depending on the initial conditions. On the other hand,
increasing $\mu$ the points in the plot form more regular paths: the effect of switching on the chemical potential is to mitigate the chaotic behavior. 

\begin{figure}[h!]
\centering
	\includegraphics[width=0.4 \textwidth]{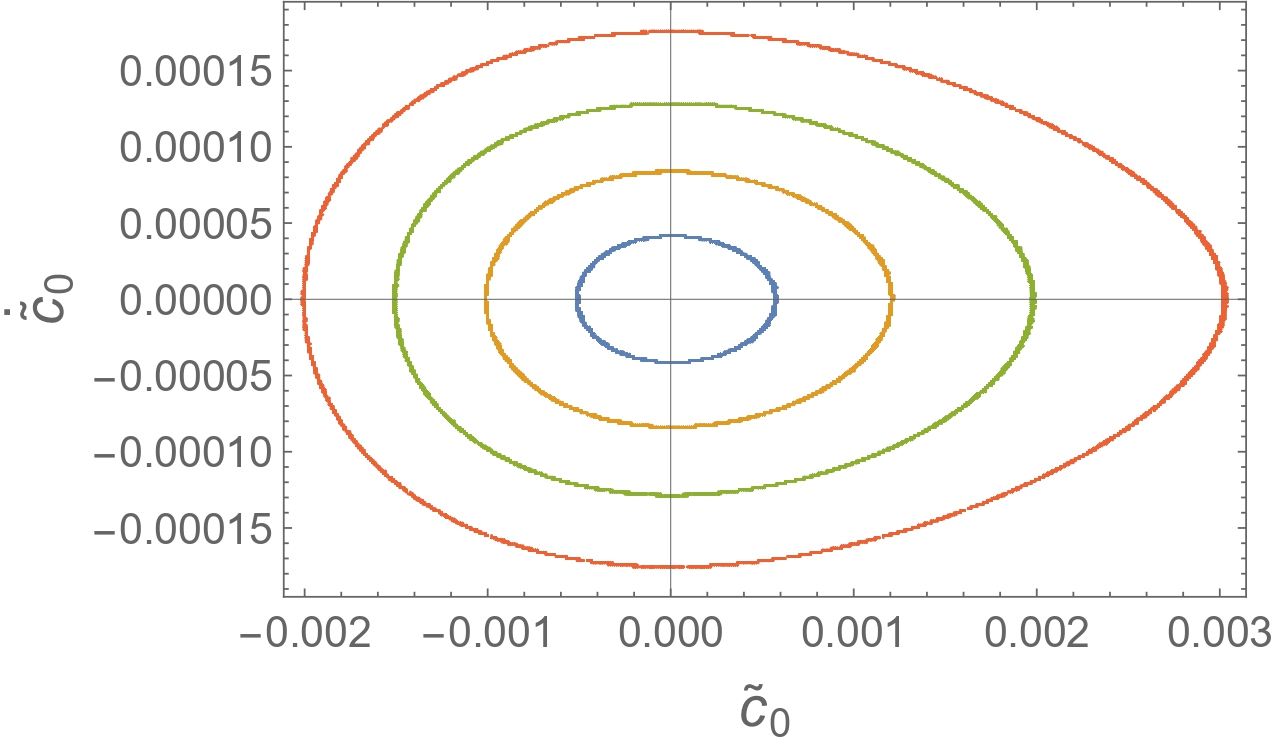} \\ \hspace*{0.3cm}
	\includegraphics[width=0.36 \textwidth]{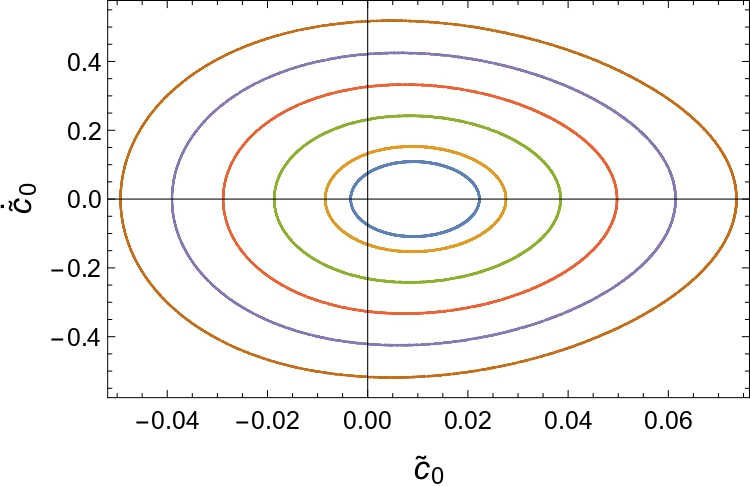}
	\caption[Poincar\'e plot for $r_0 = 1.1$ and $\mu = 1.2$]{\baselineskip 12 pt \small Poincar\'e section  in the case  $r_0 = 1.1$, $\mu = 1.2$,  energy $E = 1\times 10^{-5}$ with  $8\times 10^3$ time steps (top panel),   $r_0 = 5$, $\mu = 0$ and   energy $E = 1\times 10^{-3}$ (bottom panel). }
\label{Fig:13}
\end{figure}

For $\mu = 1.2$ and $r_0 = 1.1$ the eigenvalue $\omega_0^2$ becomes positive and the orbits form  tori, as one can see in Fig.~\ref{Fig:13}.
\noindent
Moving further away from the horizon, the Poincar\'e plots for the string dynamics  show  regular orbits regardless of $\mu$.

The Lyapunov exponents in the four dimensional $c_0$, $c_1$ phase space can be computed for the different values of $r_0$ and $\mu$ using the numerical method in \cite{sandri1996numerical},   briefly described in Appendix \ref{app:A}.  The results are shown  in Figs.~\ref{Fig:15} and \ref{Fig:16}.
\begin{figure}[h!]
\centering
	\includegraphics[width=0.4 \textwidth]{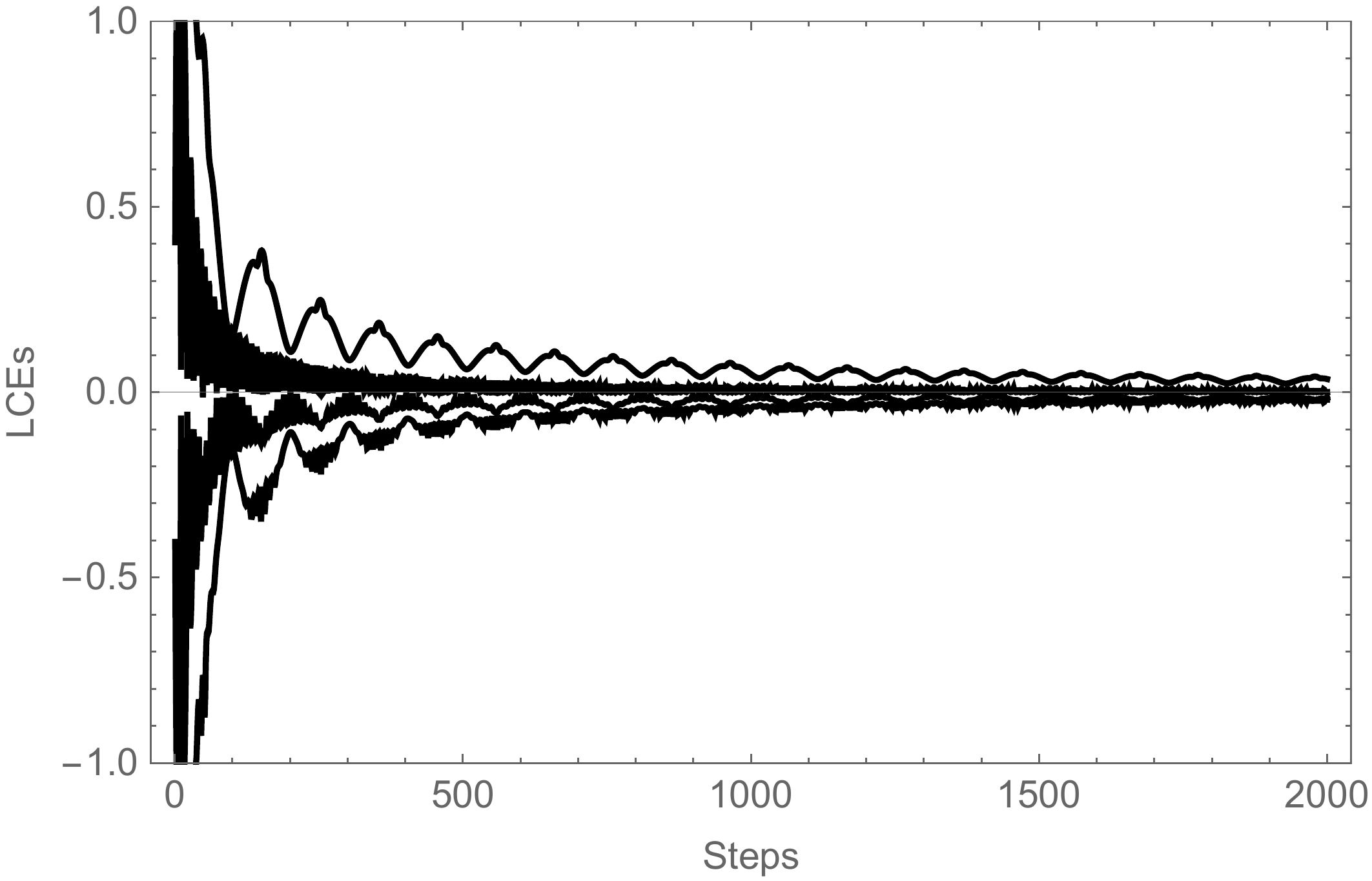}
	\includegraphics[width=0.4 \textwidth]{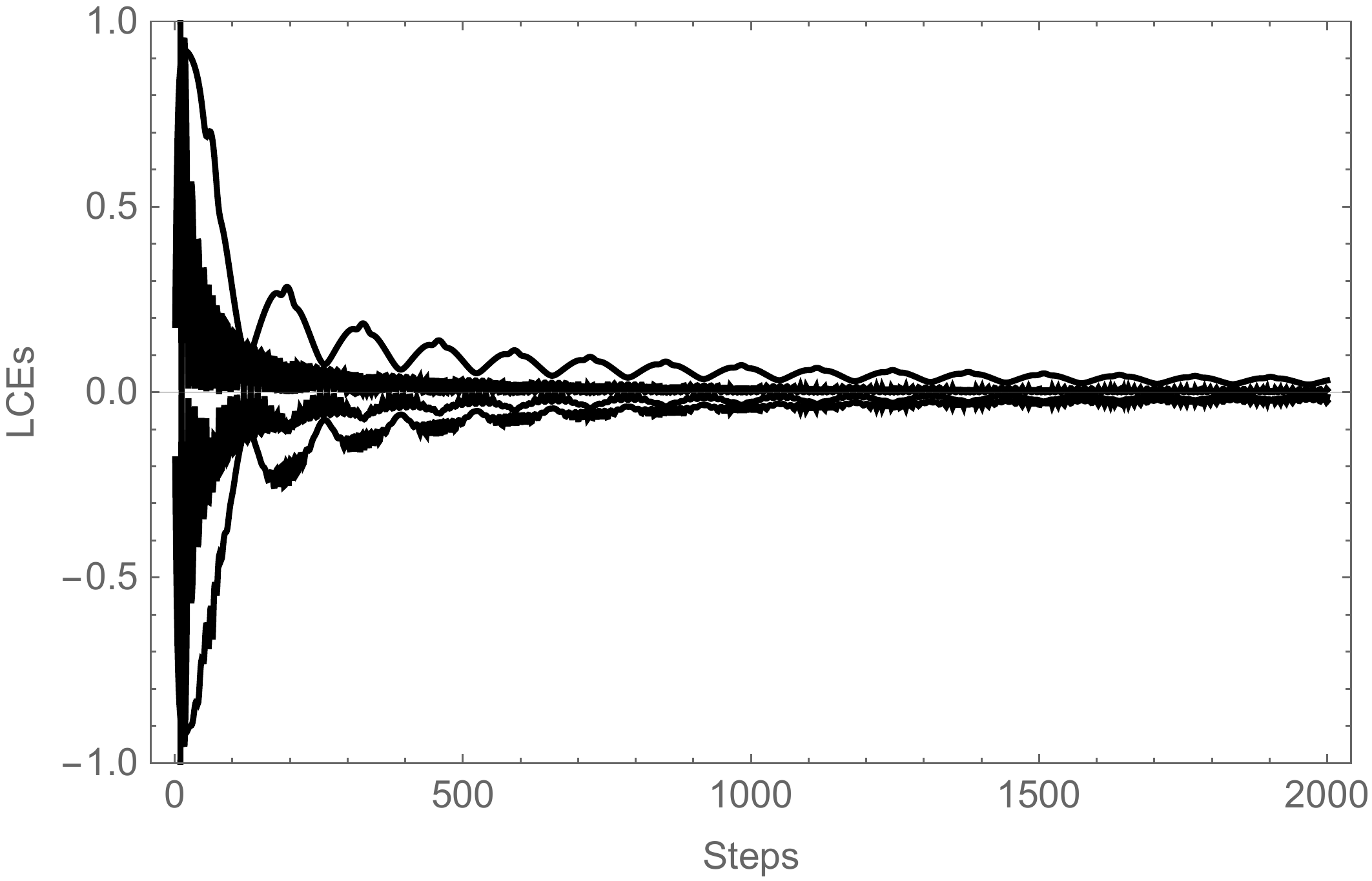}
	\caption[Convergency plots of the Lyapunov exponents for a string with $r_0=1.1$]{\baselineskip 12 pt \small Convergency plots of the four Lyapunov exponents (LCEs) in the case of a string with $r_0=1.1$,  $\mu=0$ (top panel) and $\mu=0.6$ (bottom panel),  and $2\times 10^3$ time steps.}
\label{Fig:15}
\end{figure}
Focusing on the system with $r_0=1.1$, we have evaluated the convergency plots of the four Lyapunov coefficients, one for each direction of the phase space, varying $\mu$ from  $\mu=0$ to  $\mu=1.2$. The cases $\mu=0$ and $\mu=0.6$ are displayed in Fig.~\ref{Fig:15}, the other cases are similar.  The largest Lyapunov exponent behaves as an exponentially decreasing oscillating function,  which can be extrapolated  to  large number of time steps as shown in Fig.~\ref{Fig:16}. 
\begin{figure}[h!]
\centering
	\includegraphics[width=0.4 \textwidth]{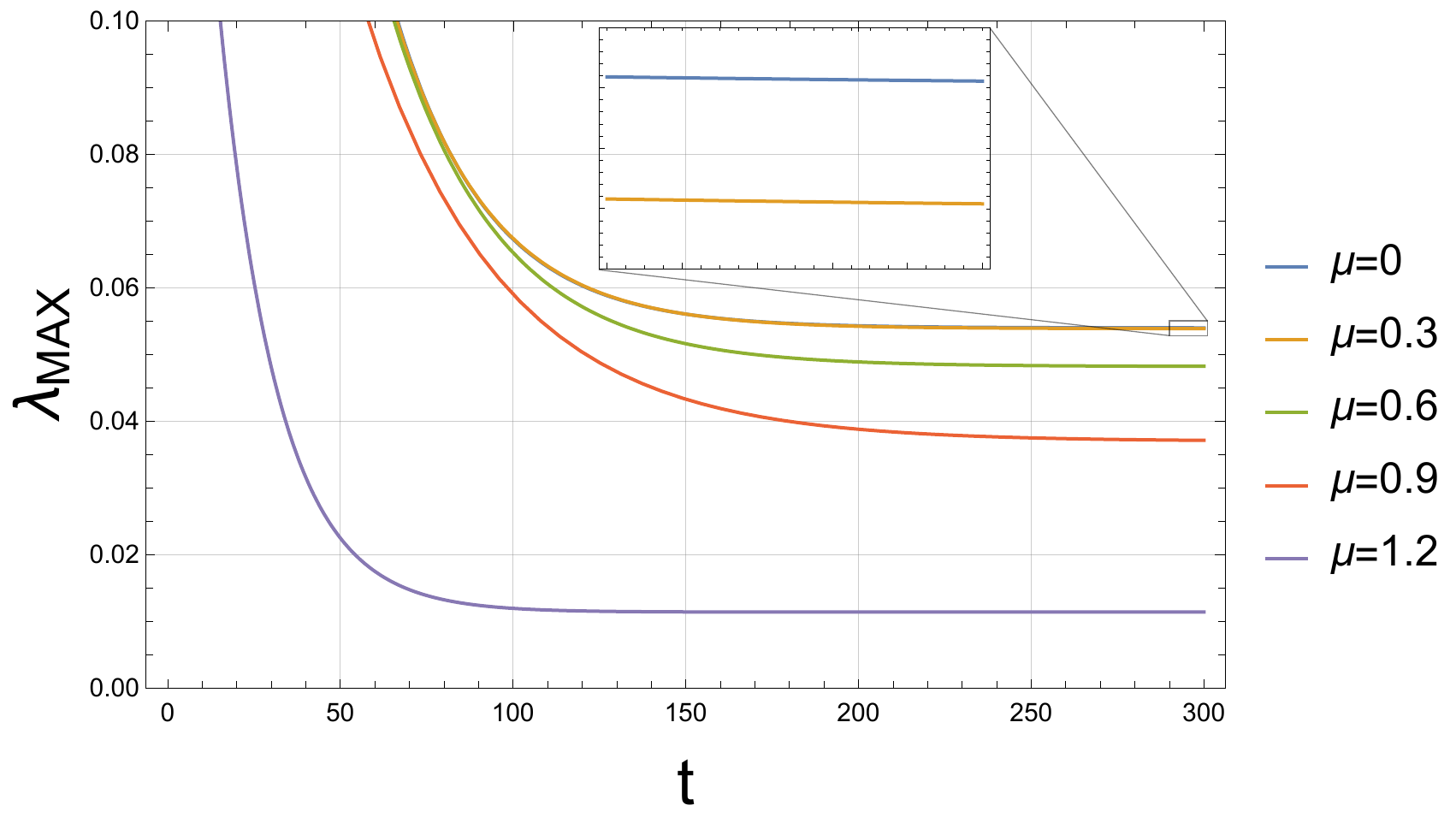}\\
	\includegraphics[width=0.42 \textwidth]{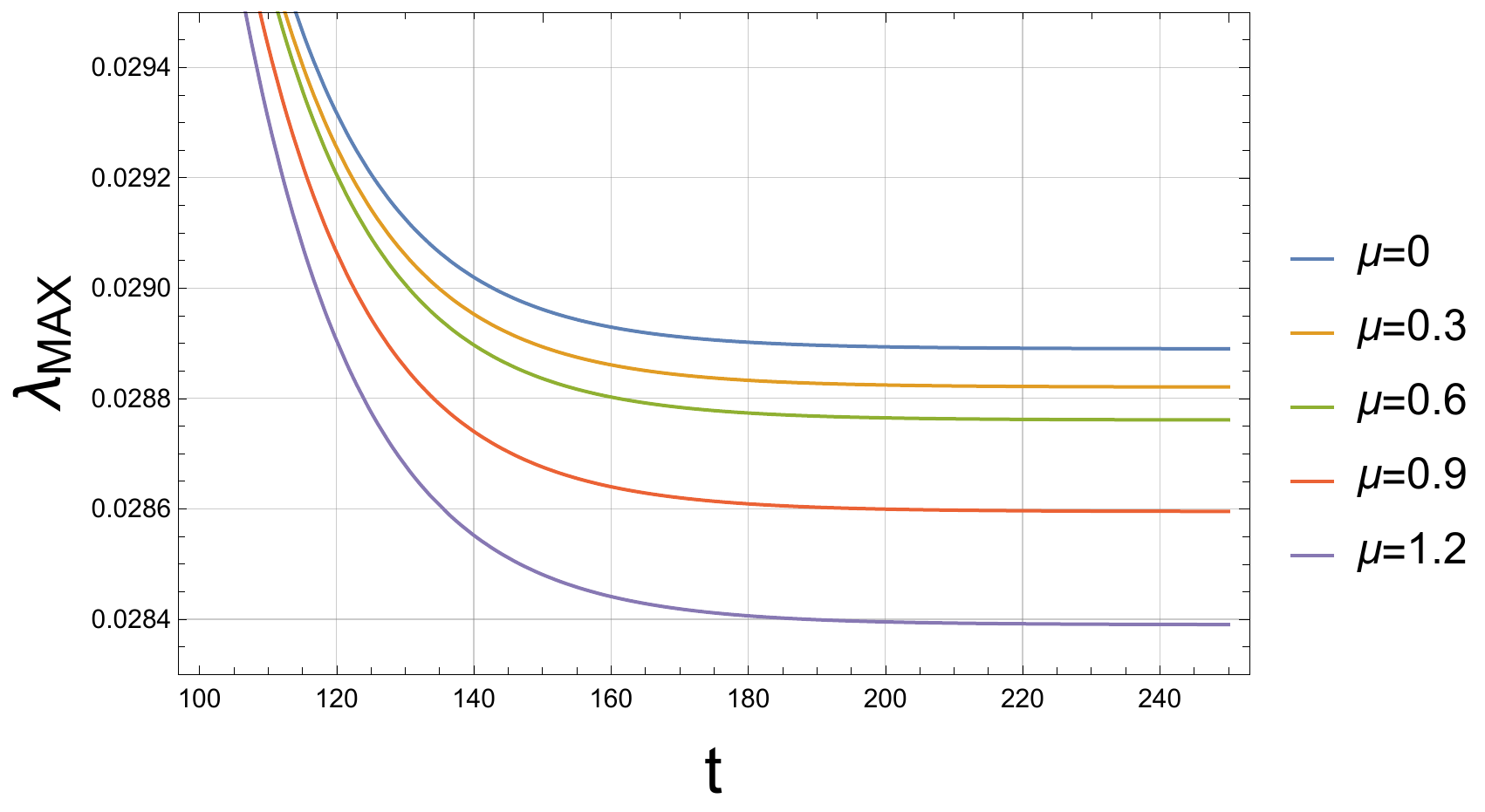}
	\caption[Fit of the convergency plot of the maximum Lyapunov coefficient for $r_0=5$]{\baselineskip 12 pt \small Fit of  the largest Lyapunov coefficient $\lambda_{MAX}$ for $r_0=1.1$ (top) and $r_0=5$ (bottom), varying $\mu$. The local maxima of plots as in Fig.~\ref{Fig:15} are fitted.}
\label{Fig:16}
\end{figure}
\noindent
 The  values resulting from the fit decrease as $\mu$ increases: the effect of the chemical potential is to soften the dependence on the initial conditions, making the string less chaotic.

To investigate the  behaviour for different $r_0$, we have computed the Lyapunov coefficients for $r_0=5$, away from the horizon,  and $\mu$ up to $\mu=1.2$. 
The convergency plots  show a rapid convergence of  all  Lyapunov coefficients  towards zero. The result of the fit for large time steps,  for different values of $\mu$ is  in the same Fig.~\ref{Fig:16}.
\begin{figure}[]
\centering
	\hspace*{0.2cm}\includegraphics[width=0.39 \textwidth]{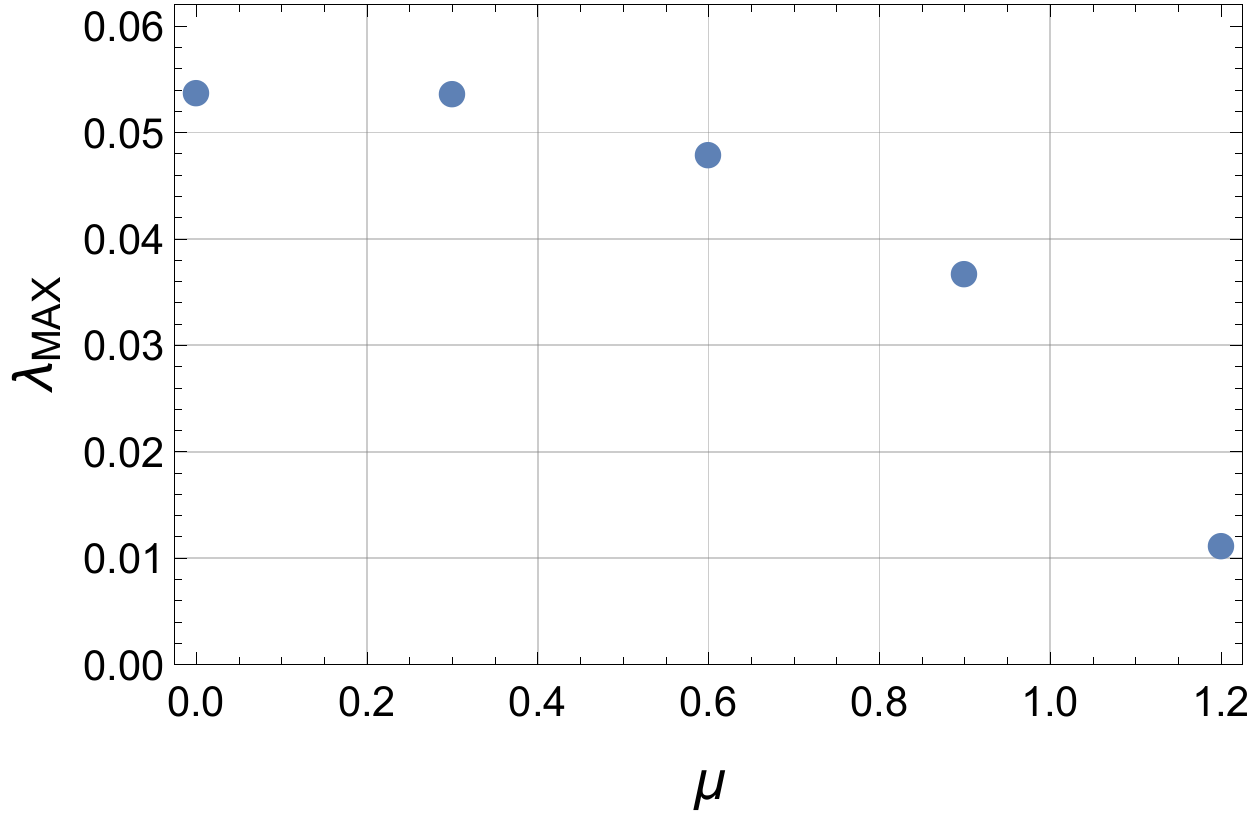}
	\includegraphics[width=0.4 \textwidth]{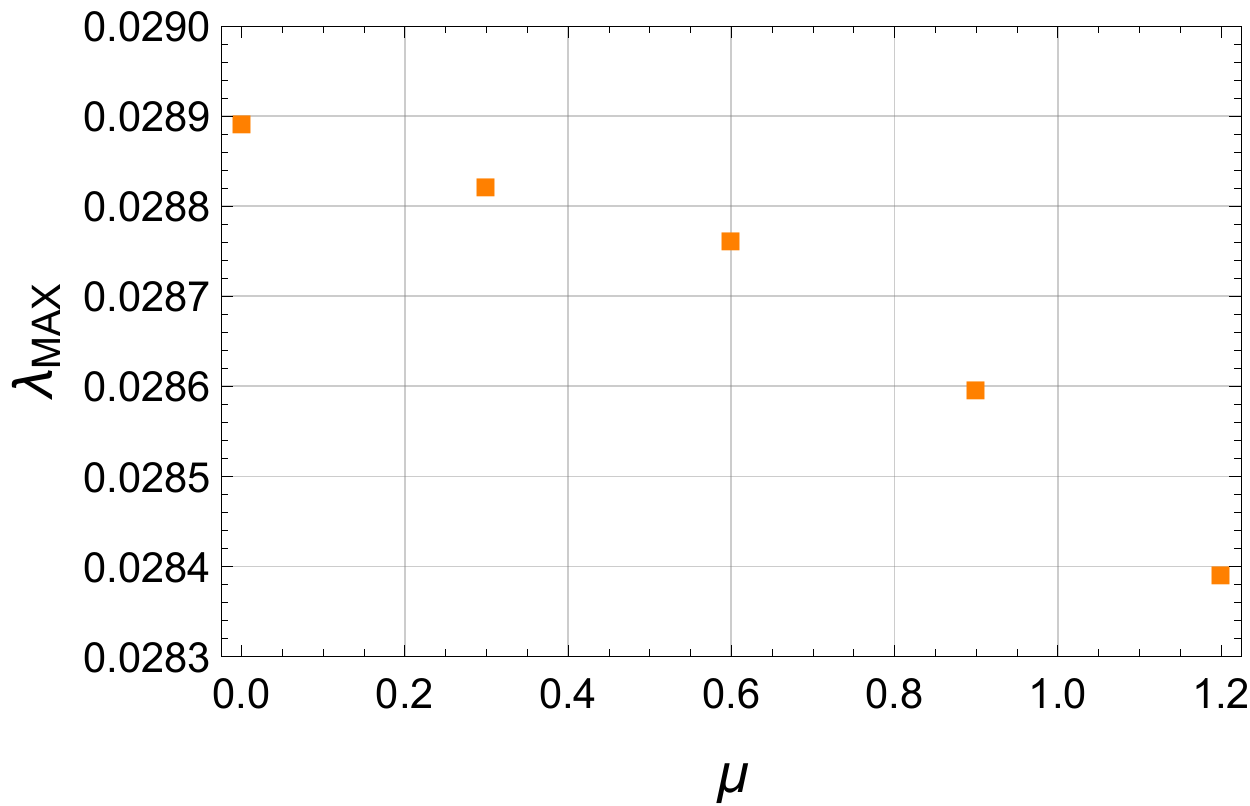}
	\caption[Largest Lyapunov exponent as a function of $\mu$]{\baselineskip 12 pt \small Largest Lyapunov exponent $\lambda_{MAX}$ vs $\mu$, for the tip position $r_0=1.1$ (top) and $r_0=5$ (bottom). }
\label{Fig:19}
\end{figure}

To summarize, the Poincar\'e plots show that chaos is produced in the proximity of the BH horizon, and that  the dynamics of the string is less chaotic as the chemical potential  increases. This is confirmed by the behavior of the  largest Lyapunov coefficient, shown in Fig.~\ref{Fig:19}.
In all cases  the bound  Eq.~\eqref{eq:1} is satisfied:  for example, for a system with $r_0 = 1.1$ and $\mu=0.6$  we have  $\lambda \simeq (2.7 \times10^{-2}) \times 2 \pi T_H$, close to the value computed for $\mu=0$  in \cite{Hashimoto:2018fkb}. There are no indications of a relaxed  bound as foreseen by Eq.~\eqref{eq:2}.
\section{Geometry with a  dilaton}
It is interesting to study a different background,  a modification of  the AdS-RN with the introduction of a warp factor,
used to implement a confining mechanism in holographic models of QCD breaking the conformal invariance \cite{Karch:2006pv}.  
The line element  is defined as

\bea
d s^2 = e^{-\frac{c^2}{r^2}} \left(-r^2  f \left( r \right)  d t^2 + r^2 d {\bar x}^2 +\frac{1}{r^2 f (r)} d r^2\right), \nn \\
\label{eq:66}  
\eea
with the same metric function $f(r)$  in Eq.~(\ref{eq:20}). The
 Hawking temperature  is in Eq.~\eqref{eq:thBH} and does not   depend on the dilaton parameter $c$. The warp factor mainly affects the IR small $r$  region, and the geometry becomes asymptotically $AdS_5$ in the UV $r \to \infty$  region. Introducing a dilaton factor has been used, in a bottom-up approach, to study  features of the QCD phenomenology at finite temperature and baryon density, namely the behaviour of the 
 quark and gluon condensates increasing  $T$ and $\mu$,  the  phase diagram,  and the in-medium broadening of the spectral functions of two-point correlators \cite{Colangelo:2010pe,Colangelo:2011sr,Colangelo:2012jy,Colangelo:2013ila}.

The analysis for a time-dependent perturbation of the static string in this background can be carried out following the  previously adopted procedure. For the square string in the background \eqref{eq:66},  the  Lyapunov exponent computed at ${\cal O}(L^2)$  reads:

\begin{equation}
\lambda=2 \pi T_H \left(1-\frac{L^2}{2} \pi T_H r_H \left(1+\frac{c^2}{r_H^2} \right) \right) \,. 
%\lambda = 2 \pi T_H - (2 \pi T_H)^2 \frac{L^2 (c^2 + r_H^2)}{4 r_H} \,.
\label{eq:67}
\end{equation}
\noindent
This expression fulfils the bound \eqref{eq:1}.

\begin{figure*}[]
\centering
\makebox[\linewidth][c]{{
	{\includegraphics[width=0.35 \textwidth]{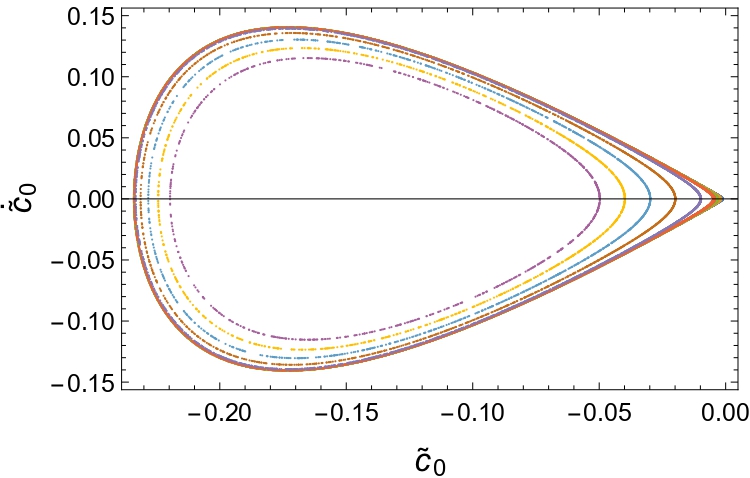} \quad
	 \includegraphics[width=0.37 \textwidth]{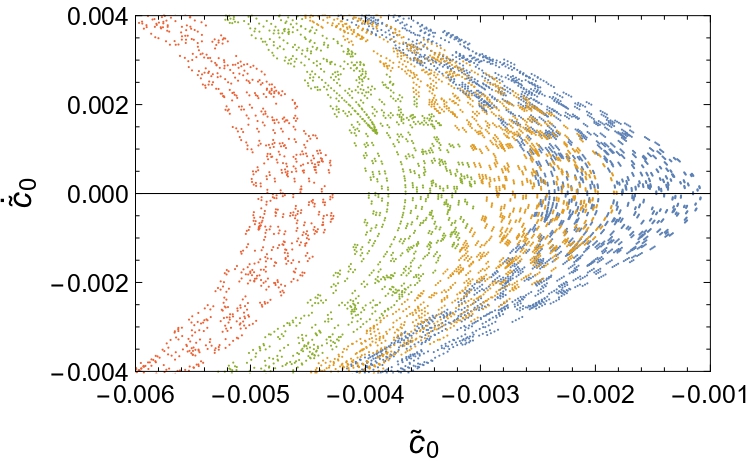}}}} \quad
\makebox[\linewidth][c]{{
	{\includegraphics[width=0.35 \textwidth]{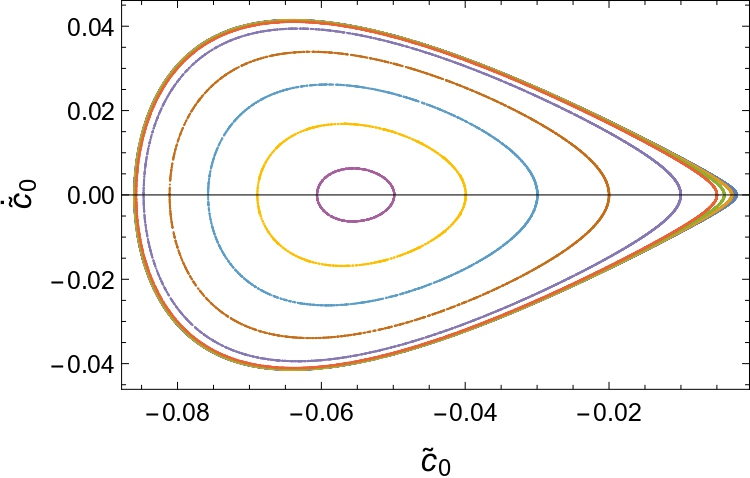} \quad
	 \includegraphics[width=0.38 \textwidth]{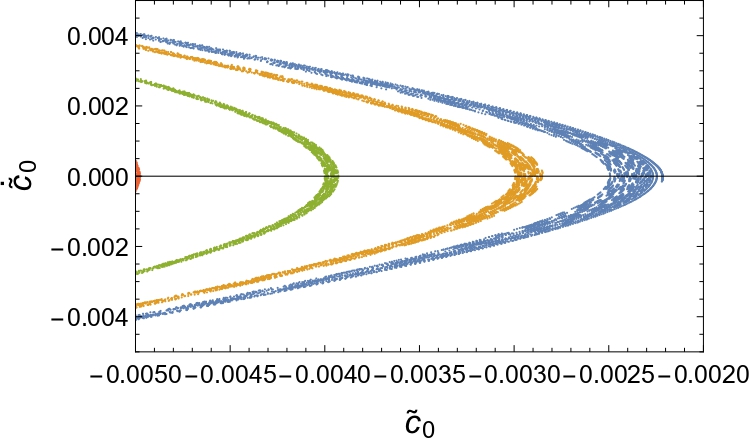}}}} \quad
\caption[Poincar\'e plots for $r_0 = 1.1$ and different value of $\mu$]{\baselineskip 12 pt \small Top: Poincar\'e section for the perturbed string in the background geometry with warp factor \eqref{eq:66},  for $r_0 = 1.1$, $\mu = 0$ and parameter of the dilaton $c = 1$,  energy  $E=1\times 10^{-5}$ and $8\times 10^3$ time steps (left plot). The right plot enlarges the left one in the small $\tilde{c}_0$, $\dot{\tilde{c}}_0$ region. The bottom panels correspond to  $c = 2$.}\label{Fig:dil}
\end{figure*}

\begin{figure}[h]
\centering
	\includegraphics[width=0.4 \textwidth]{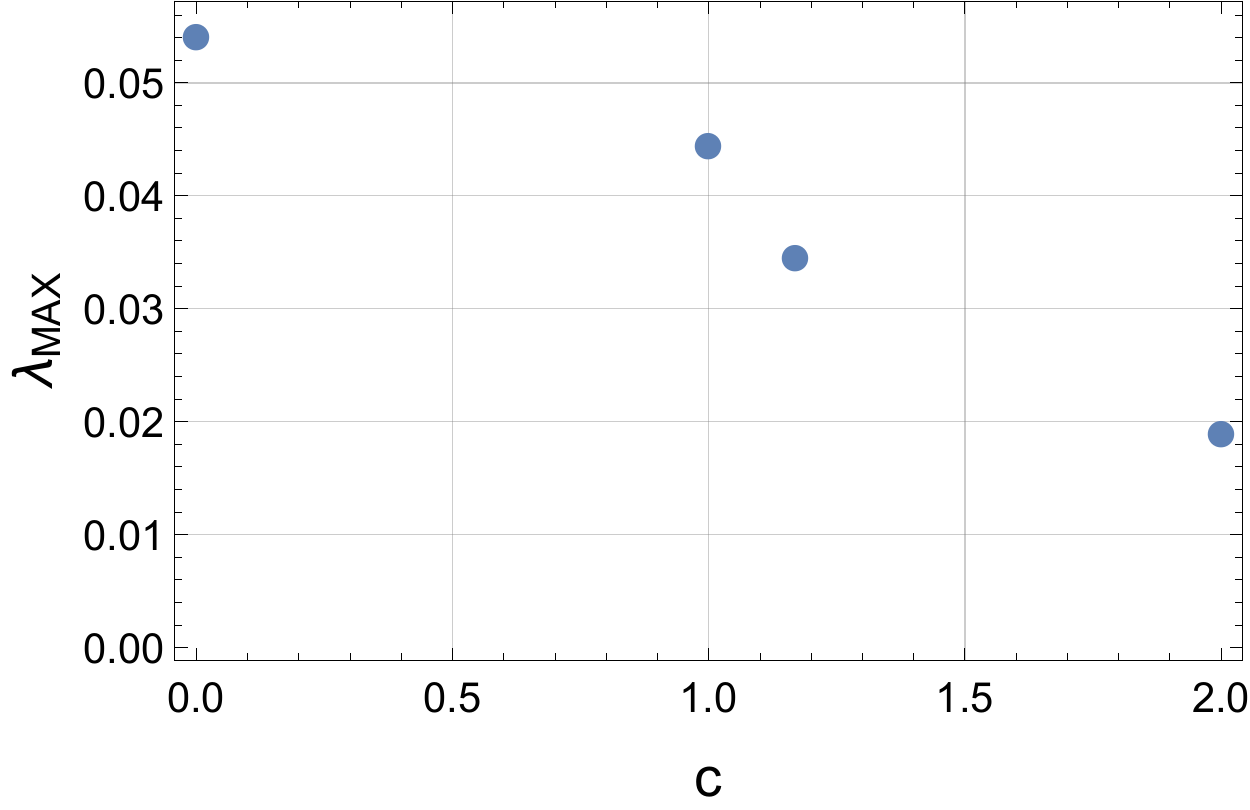}
	\caption[Largest Lyapunov exponent as a function of $c$]{\baselineskip 12 pt \small Largest Lyapunov exponent for $r_0=1.1$ and $\mu = 0$, increasing the dilaton constant $c$.}
\label{Fig:22}
\end{figure}

To study the dependence of chaos on the dilaton parameter $c$,  we inspect the Poincar\`e plots and compute the Lyapunov  exponents.  
The Poincar\`e section for  $r_H = 1$, $r_0 = 1.1$, $\mu = 0$ and $c = 1$  is shown in Fig.~\ref{Fig:dil}.
For small values of $\tilde c_0$, $\dot {\tilde c}_0$  the section shows patterns hinting for a less chaotic system as the  constant $c$ increases. This is confirmed  by  the Poincar\`e plot for $c = 2$, which shows regular orbits also in the phase space region of small  $\tilde c_0$ and $\dot {\tilde c}_0$.
Therefore, increasing the dilaton parameter $c$ the system is less chaotic. It can also be inferred from  Fig.~\ref{Fig:22}, where
 the Lyapunov coefficient for the string with $r_H = 1$, $r_0 = 1.1$, $\mu = 0$ and a few values of $c$ is drawn: the exponent monotonically decreases  vs $c$.
\\
\\

\section{Conclusions}
The investigation of a holographic dual of the heavy quark-antiquark system  confirms the bounds \eqref{eq:1}  also in the case of finite baryonic chemical potential.
With increasing $\mu$ the system is less chaotic. This  agrees with the conclusion obtained considering the charged particle motion in the RN AdS background, 
for which a reduction of the chaotic behaviour is observed increasing the chemical potential \cite{Ageev:2018msv}.
Decrease in chaoticity is also observed for a thermal background involving a dilaton warp factor.   

 Even though our study is limited to  small perturbations of the static string configuration,  it seems unlikely that  the analysis of  large fluctuations  would lead to different results:  In the case $\mu=0$ the numerical computation of large string fluctuations  around the static configuration confirmed the results obtained for small perturbations  \cite{Hashimoto:2018fkb}. 
This induces us to conclude  that the  bound \eqref{eq:1} continues to hold in the case of finite chemical potential.

A  possible extension of  our analysis concerns the interplay between chaos and time-dependent  background geometry, namely 
 the hydrodynamic metric  worked out in \cite{Chesler:2008hg,Chesler:2009cy,Bellantuono:2015hxa, Bellantuono:2017msk}.  It would be interesting to establish the existence of a bound analogous to Eq.~\eqref{eq:1} also in these cases.
  
\vspace*{0.5cm}
\noindent {\bf Acknowledgements.} 
We thank F. Giannuzzi, A. Mirizzi  and S. Nicotri for discussions. 
This study has been  carried out within the INFN project (Iniziativa Specifica) QFT-HEP.

\appendix
\numberwithin{equation}{section}
\section{Computation of the Lyapunov exponents}\label{app:A}

To compute the Lyapunov exponents we use a method that can be applied to any n-dimensional dynamical system defined by the  equation 

\bea
\dot{x} = F(x)
\label{eq:A1}
\eea 
where $\dot{x} = d x /d t$ \cite{sandri1996numerical}. In our case we have a 4-d Hamiltonian dynamical system, with the Hamilton equations obtained from the Legendre transformation of Eq.~(\ref{eq:64}). The point $x(t)$  in the phase space is represented by the  variables $\tilde{c}_0(t)$, $\tilde{c}_1(t)$ and their conjugates momenta.
The Lyapunov coefficients, describing the exponential rate growth of the distance between two initially near trajectories, are given by

\bea
\lambda(x_0,u_0) &=& \lim_{t \rightarrow \infty} \frac{1}{t} \ln \frac{||u_t||}{||u_0||} \nn \\
&=& \lim_{t \rightarrow \infty} \frac{1}{t} \ln || D_{x_0}f^t (x_0)\cdot u_0||.
\label{eq:A2}
\eea
\noindent
In \eqref{eq:A2} 
 $||u_0||$ is the length of the vector representing the initial perturbation between two near trajectories,  $u_t$ is its evolution at time $t$, and the second equality is obtained from
the truncation
\begin{equation}
u_t = f^t (x_0 + u_0) - f^t (x_0) 
= D_{x_0}f^t (x_0)\cdot u_0,
\label{eq:A3}
\end{equation}
\noindent
where $f^t(x_0)$ is the solution of Eq.~(\ref{eq:A1}) with initial condition $x_0$. This vector satisfies the so-called variational equation:

\bea
\dot{\Phi}_t (x_0) &=& D_x F(f^t (x_0)) \cdot \Phi_t (x_0), \nn \\ 
\Phi_0(x_0) &=& I,
\label{VarEq}
\eea
\noindent
where $\Phi_t (x_0) = D_{x_0} f^t (x_0)$. 

 To compute the Lyapunov exponents  both Eqs.~(\ref{eq:A1}) and (\ref{VarEq}) must be solved, namely using  a Runge-Kutta method 
 fixing a  time step size  $s$  and iterated $K$ times in the time interval $T$.

From  Eq.~(\ref{eq:A2}) the largest Lyapunov coefficient (denoted as LCE of order $1$) is obtained. It is useful to generalize the definition for LCEs of order $p$, describing  the mean rate growth of a $p$-dimensional volume in the tangent space to the trajectory. They are defined by

\bea
\lambda^p(x_0,U_0) = \lim_{t \rightarrow \infty} \frac{1}{t} \ln ||\text{Vol}^p (D_{x_0}f^t (U_0))||, \,\,\,\,
\label{eq:A5}
\eea
\noindent
where $U_0$ is an initial parallelepiped identified by the initial conditions of the near $p+1$ trajectories. It is always  possible to find $p$ linearly independent vectors such that 

\bea
\lambda^p(x_0,U_0) = \lambda_1 +\lambda_2 + \dots + \lambda_p.
\label{eq:A6}
\eea 
\noindent
Therefore, each LCE of order $p$ is given by the sum of the $p$ largest LCEs of order 1. For $p = n$ we obtain the mean exponential rate of growth of the phase space volume,  given by the sum of the whole spectrum of LCEs.  This property can be used to implement an algorithm to evaluate convergency plots of the spectrum of the Lyapunov exponents. The algorithm makes use of the Gram-Schmidt procedure to generate a set of orthonormal vectors. Given an $n$-dimensional solid $U_0$ identified by $n$-vectors $\{ u_1, \dots, u_n \}$ we have

\be
\text{Vol}\{ u_1, \dots, u_n \} = ||w_1|| \dots|| w_n ||,
\label{eq:A7}
\ee
where the $w$ vectors are the orthonormal vectors obtained by the Gram-Schmidt procedure on the $u$ vectors. Hence, starting from an initial condition $x_0$ in the phase space and an $n \times n$ matrix, that is the initial condition $U_0 = \{ u^{0}_1, \dots, u^{0}_n \}$ for Eq.~(\ref{VarEq}), we integrate the system of equations \eqref{eq:A1} and \eqref{VarEq}. After each iteration,  the evolution of the tangent vectors is obtained: $U_1$ for the first iteration, and so on. The new vectors must be orthogonalized at each iteration. During the $k$-th step the $n$-dimensional volume increase by a factor $||w^k_1|| \dots|| w^k_n ||$, where $\{ w^k_1, \dots, w^k_n \}$ is the set of orthogonal vectors calculated from $U_k$. From Eq.~(\ref{eq:A5})  we have for $p = n$:

\bea
\lambda^n(x_0,U_0) = \lim_{k \rightarrow \infty} \frac{1}{k T} \sum_{i=1}^k\ln (||w^i_1|| \dots|| w^i_n ||). \,\,\,\,\,\,\,\,\,\,\,
\label{eq:A8}
\eea
Subtracting $\lambda^{n-1}$ and using the property in Eq.~(\ref{eq:A6}), we obtain the $n$-th LCE of order $1$:

\bea
\lambda_n = \lim_{k \rightarrow \infty} \frac{1}{k T} \sum_{i=1}^k\ln ||w^i_n ||.
\label{eq:A9}
\eea
The procedure allows to compute the whole spectrum of the Lyapunov exponents for the total number of steps $K$ reasonably large:
\begin{equation}
\begin{aligned}
\lambda_1 &\sim \frac{1}{K T} \sum_{i=1}^K\ln ||w^i_1 ||, \\  & \dots, \\ \lambda_n &\sim \frac{1}{K T} \sum_{i=1}^K\ln ||w^i_n ||.
\end{aligned}
\label{eq:A10}
\end{equation}
%
%\newpage
\bibliographystyle{apsrev4-1}
\bibliography{biblio}
\end{document}